\documentclass[showpacs, pra,twocolumn,preprintnumbers ,amsmath, amssymb, superscriptaddress, aps]{revtex4-2}
\usepackage{amsfonts}
\usepackage{color}
\usepackage{amsmath,amssymb}
\usepackage{pifont}
\usepackage{amssymb}  
\usepackage{bbold}
\usepackage{float}
\usepackage{subfloat}

\usepackage[caption=false]{subfig}
\usepackage{tikz}
\usepackage{makecell}
\usepackage{subfig}
\usepackage{pifont}   
\usepackage{graphicx} 
\graphicspath{{Figures/}}
\usepackage{dcolumn}  
\usepackage{bm}       
\usepackage{multirow} 
\usepackage{placeins}
\usepackage[colorlinks]{hyperref}
\usepackage{mathtools}
\usepackage{appendix}

\def \be{\begin{align}}
	\def \ee{\end{align}}
\def \bea{\begin{eqnarray}}
	\def \eea{\end{eqnarray}}

\begin{document}

	\title{
		{Energy levels of gapped graphene quantum dots in external fields}}
	\date{\today}
	\author{Ahmed Bouhlal}
	\email{bouhlal.a@ucd.ac.ma}
	\affiliation{Laboratory of Theoretical Physics, Faculty of Sciences, Choua\"ib Doukkali University, PO Box 20, 24000 El Jadida, Morocco}
	\author{Mohammed El Azar}
	\affiliation{Laboratory of Theoretical Physics, Faculty of Sciences, Choua\"ib Doukkali University, PO Box 20, 24000 El Jadida, Morocco}
	\author{Ahmed Siari}
	\affiliation{Laboratory of Theoretical Physics, Faculty of Sciences, Choua\"ib Doukkali University, PO Box 20, 24000 El Jadida, Morocco}
	\author{Ahmed Jellal}
	\email{a.jellal@ucd.ac.ma}
	\affiliation{Laboratory of Theoretical Physics, Faculty of Sciences, Choua\"ib Doukkali University, PO Box 20, 24000 El Jadida, Morocco}
	\affiliation{Canadian Quantum Research Center, 204-3002 32 Ave Vernon,  BC V1T 2L7, Canada}

	\pacs{}

\begin{abstract}

We investigate the energy levels of fermions within a circular graphene quantum dot (GQD) subjected to external magnetic and Aharonov-Bohm fields. Solving the eigenvalue equation for two distinct regions allows us to determine the eigenspinors for the valleys $K$ and $K^\prime$. By establishing the continuity of eigenspinors at the GQD interface, we derive an equation that reveals the reliance of energy levels on external physical parameters. Our observations suggest that the symmetry of energy levels hinges on the selected physical parameters. We observe that at low magnetic fields, the energy levels display degeneracy, which diminishes as the field strength increases, coinciding with the convergence of energy levels toward the Landau levels. We illustrate that the introduction of a magnetic flux into the GQD leads to the creation of an energy gap, extending the trapping time of electrons without perturbing the system. Conversely, the addition of gap energy widens the band gap, disrupting the system's symmetry by introducing new energy levels.

\vspace{0.25cm}
\noindent PACS numbers:  73.22.Pr, 72.80.Vp, 73.63.-b\\
	\noindent Keywords: Graphene, magnetic field, quantum dot, magnetic flux, energy levels, energy gap, Landau levels.

\end{abstract}
\maketitle

\section{Introduction}
Carbon-based electronics are a rapidly developing technology with great potential. Since its isolation in 2004, graphene has become the leading candidate to replace silicon in solid-state physics \cite{novoselov2004electric, geim2007rise, xia2010index}. It is a one-atom-thick allotrope of carbon. It is arranged in a honeycomb lattice with $sp^2$ hybridization. It has remarkable properties: it is the strongest material with record thermal conductivity and high electron mobility \cite{geim2007rise, hwang2008acoustic}. Since then, the number of publications has exponentially grown, driven by promising theoretical predictions of exceptional electronic and mechanical properties. The unique band structure of its hexagonal carbon lattice gives graphene its electronic properties. In the Brillouin zone, the valence band and the conduction band intersect at two points, $K$ and $K^\prime$, giving rise to a cone-shaped spectrum \cite{sun2008ultrafast, kim2009large}. As the electrons traverse through the carbon lattice, they behave like ultra-relativistic particles, moving at a speed about 300 or so times slower than the speed of light \cite{neto2009electronic}. The Dirac Hamiltonian, which governs the evolution of graphene's electron wave functions, describes the behavior of relativistic fermions \cite{geim2009graphene}. Graphene has highly specific transport properties. It has a linear dispersion relationship, unlike conventional semimetals, which have a parabolic relationship. These unique properties provide the potential for the development of graphene-based electronic systems that can process information at speeds 10 times faster than current systems. Graphene, while having many promising properties, lacks the energy gap that is necessary for the production of transistors. This absence of a band gap impedes the creation of logic gates and results in the production of low-quality devices with poor signal modulation and high leakage current \cite{geim2007rise, schwierz2010graphene}.

One strategy for opening a band gap in graphene is to cut it into nanoribbons, which create such a band by confining the wave function \cite{han2007energy}. One approach involves using graphene quantum dots, which are 0D structures, along with 1D graphene nanoribbons to reduce the dimensionality of the 2D system. In addition, a band gap \cite{guttinger2009electron, libisch2010transition} can be created by isolating a graphene quantum dot (GQD).Also known as artificial atoms, GQDs share properties with atoms and are made of very small pieces of graphene \cite{zhu2012control, li2013focusing}. Their potential for future quantum information applications \cite{li2013focusing}. Researchers have been trying to trap electrons in graphene-based quantum dots (GQDs) for a variety of applications, including electronic circuits, gas detection \cite{sun2013recent}, and photovoltaic systems \cite{bacon2014graphene}. The Klein tunneling effect \cite{katsnelson2006chiral} prevents carriers from being confined, making graphene-based quantum structures challenging to fabricate for electronic devices. Manipulation of the energy of carrier states in graphene can be achieved by using external magnetic fields, leading to the appearance of Landau levels in an infinite graphene sheet, or by using finite-size GQDs \cite{ponomarenko2008chaotic}. The phenomenon of confinement induced by a magnetic field in graphene quantum dots has been studied in detail in a number of references \cite{schnez2008analytic, recher2009bound, orozco2019enhancing}.

Several techniques have been proposed to trap electrons in graphene and create intriguing graphene-based structures.~Research has demonstrated that Dirac fermions in GQDs can be confined through the use of external magnetic fields \cite{giavaras2009magnetic}, finite-mass terms with electrostatic potentials \cite{recher2009bound}, or infinite-mass terms \cite{thomsen2017analytical, orozco2019enhancing}. An experimental study was carried out to confine electrons in an electrostatically confined monolayer GQD with orbits and valleys \cite{freitag2016electrostatically}. Moreover, the energy states of circular graphene quantum dots (GQDs) have been explored under the influence of an external magnetic field perpendicular to the graphene plane \cite{Myoung19, farsi2020energy}. Additionally, investigations have been conducted considering two mass terms, with one situated inside and the other outside the quantum dot \cite{belokda2022energy}. These theoretical studies have shown that energy levels exhibit symmetric and anti-symmetric behaviors under appropriate conditions of physical parameters characterizing system-based graphene. Furthermore, it was found that the introduction of an energy gap diminishes the electron density within the quantum dot and alters its electronic characteristics, resulting in the transient confinement of electrons.

We investigate the electron confinement energy levels within a circular graphene quantum dot (GQD) of radius $R$ exposed to an external magnetic field applied in the surrounding region, along with a mass term (energy gap $\Delta$) and an Aharonov-Bohm field (magnetic flux $\phi_i$) applied internally. Our study relies on solving the Dirac equation to find eigenspinors in two regions separated by the radius  $R$. Using the continuity of the eigenspinors at  boundary condition, we derive an equation that describes the relation between the energy levels and the physical parameters that characterize our system. Due to the difficulty of obtaining an explicit solution for the energy, we opt for a numerical approach to highlight the fundamental features of our system. For this, 
 we study the dependence of the energy levels on the angular momentum $m$, the radius of the GQD $R$, the magnetic field $B$, the energy gap $\Delta$, and the magnetic flux $\phi_i$. 
 As a consequence, we demonstrate that the inclusion of $\phi_i$ expands the band gap, creating a separation between the valence and conduction bands without disrupting the symmetry of the energy levels. This extension enhances the trapping time of electrons within the quantum dot. Additionally, the introduction of an energy gap also widens the band and allows new energy levels to appear, but it breaks the symmetry of energy. In conclusion, we assert that magnetic flux serves as a tunable parameter capable of controlling the electronic properties of our system.

The present paper is structured as follows. The Hamiltonian is presented, and the spinors and energy spectra in the two regions are determined in Sec. \ref{MH}. Using the boundary continuity conditions, we obtain an equation that gives the dependence of the energy levels on the parameters of our system. In Sec. \ref{results} the numerical results of our analysis are well presented, and we give an interpretation of the results found. Finally, Sec. \ref{concl} presents the conclusions of this work.

\section{MODEL HAMILTONIAN}\label{MH}
\begin{figure}[ht]
	\centering
	\includegraphics[scale=0.5]{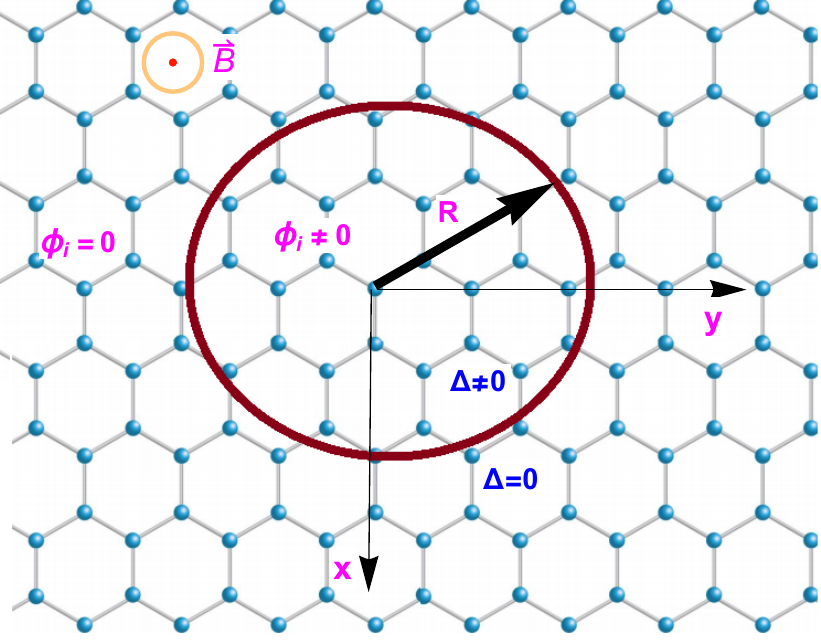}
	\caption{ (color online) Illustration of a GQD of radius $R$ surrounded by a graphene sheet in the presence of a perpendicular magnetic field $B$, Aharonov-Bohm flux fields $\phi_i$ and gap energy $\Delta$. }\label{system}
\end{figure}

To initiate our study, we will begin by describing the system at hand, which is represented as a GQD, with $R$ denoting the radius as depicted in  Fig. \ref{system}. 
Taking into account the magnetic and Aharonov-Bohm flux fields along the $z$-direction defined by $\vec{B} = B(r) \vec{e}_z$, our system can be modeled as a circularly symmetric quantum dot (QD) with
\begin{equation}
B(r) = \begin{cases}0, & r<R \\B, & r \geq R \end{cases}.
\end{equation}
Both fields produce the vector potential   $ \vec   {A}= \vec{A_1} + \vec{A_2}$ \cite{ikhdair2015nonrelativistic} where $\vec{A_1}=F(r) \vec{e_\theta}$ and $\vec{A_2}= \frac{\phi_{AB}}{2 \pi r} \vec{e_\theta}$ with
\begin{equation}
F(r)= \begin{cases}\frac{B}{2 r}\left(r^2-R^2\right), & r\geq R \\ 0, & r<R \end{cases}, \qquad  \phi_i=\frac{\phi_{AB}}{\phi_0}
\end{equation}
knowing that magnetic flux $\phi_i$ is measured in flux quantum units $\phi_0=h/e$. 

We model our system by the following Hamiltonian near the Dirac points
\begin{equation}\label{hamilt}
H_s=v_F\left(\pi_x \sigma_x+ s \pi_y \sigma_y\right)+s \Delta(r) \sigma_z
\end{equation}
with $v_F=10^6 \mathrm{~m} / \mathrm{s}$ is the Fermi velocity, $s= 1 (-1)$ refers to the valley $K$ and $K^{\prime}$, $\pi_i=p_i+e A_i$ is the conjugate momentum, $\sigma_i (i=x, y, z)$ are the Pauli matrices in the basis of the two sublattices of the $A$ and $B$ atoms, and $\delta(r)$ is the energy gap with
\begin{equation}
	\Delta (r) = \begin{cases}
		\hbar v_F \delta, & r<R\\ 0, & r \geq R.
	\end{cases}
\end{equation}
The geometry of our system, as depicted in Fig. \ref{system}, necessitates operating in polar coordinates $(r, \theta)$. Consequently, the corresponding Hamiltonian \eqref{hamilt}  transforms into
\begin{widetext}	
\begin{equation}\label{hams}
	H_s=\hbar v_F\begin{pmatrix}
		s \delta & e^{- i s \theta}\left[-i \frac{\partial}{\partial r}-i s \left(-\frac{i}{r} \frac{\partial}{\partial \theta}+\frac{e A \theta}{\hbar}\right)\right] \\
		e^{i  s \theta}\left[-i \frac{\partial}{\partial r}+i s\left(-\frac{i}{r} \frac{\partial}{\partial \theta}+\frac{e A \theta}{\hbar}\right)\right] & -s \delta
	\end{pmatrix}.
\end{equation}
\end{widetext}
As the Hamiltonian \eqref{hams} commutes with the total angular momentum $J_z=L_z+\hbar \sigma_z / 2$, we can represent the spinors, which are simultaneous eigenvectors of both $H_s$ and $J_z$, in the following manner
\begin{equation}
	\Psi_s(r, \theta)=e^{i m \theta}\dbinom{
		\Phi_A(r)} 
		{i e^{i s \theta} \Phi_B(r)}
\end{equation}
where the quantum numbers  $m = 0, \pm 1, \pm 2\cdots $ represent the eigenvalues of $J_z$.

To find the solutions for the energy spectrum, we solve the eigenvalue equation \(H_s \Psi_s(r, \theta)=E_s \Psi_s(r, \theta)\), leading to the emergence of two coupled equations 
\begin{align} \label{eq7}
&  \left[ \frac{\partial}{\partial r} + \frac{s}{r}\left(m  + s\right) +\frac{s e A_\theta}{\hbar } \right] \Phi_{B, s}=(\epsilon-s \delta) \Phi_{A, s} \\\label{eq8}
&  \left[ - \frac{\partial}{\partial r} + \frac{m s}{r}  +\frac{s e A_\theta}{\hbar } \right] \Phi_{A, s}=(\epsilon+s \delta) \Phi_{B, s}
\end{align}
where we have introduced the rescaled energy $\epsilon = \frac{E_s}{\hbar v_F}$. Now, by inserting the Eq. \eqref{eq8} into Eq. \eqref{eq7} we find a second degree differential equation for $\Phi_{A, s}(r)$.
\begin{equation} \label{eqdif}
	\left[\frac{\partial^2}{\partial r^2} + \frac{1}{r} \frac{\partial}{\partial r}-\frac{j^2}{r^2} - \frac{r^2}{4 l_B^4} - \frac{j + s}{l_B^2}+k^2\right] \Phi_{A, s} =0
\end{equation}
where we have set the parameter $j = m - \alpha + \phi_i$, $\alpha = \frac{R^2}{2 l_B^2}$, $l_B=\sqrt{\frac{\hbar}{e B}}$ is the magnetic length, and  the wave number  $k= \sqrt{\epsilon^2 - \delta^2}$. To address Eq. \eqref{eqdif}, we examine each region independently. For $r<R$ within the quantum dot, Eq. \eqref{eqdif} simplifies to
 \begin{equation} \label{eqdif1}
 	\left[\rho^2 \frac{\partial^2}{\partial \rho^2} + \rho  \frac{\partial}{\partial \rho} + \rho^2-l^2 \right] \Phi_{A, s} =0
 \end{equation}
with the variable change as $\rho = k r$. 
The general solution of Eq. \eqref{eqdif1} associated with the quantum number $l = m  + \phi_i$ involves the Bessel functions $J_l(\rho)$. The second spinor component can be obtained from Eq. \eqref{eq8} and then  combining everything to get the  eigenspinors $\Psi_{in, s}(r, \theta)$
\begin{widetext}
	\begin{equation}
	\Psi_{in, s}= C_1 e^{i\left(l - \phi_i\right) \theta}\begin{pmatrix}
		J_l(k r)\\ 
		{i e^{i s \theta} \sqrt{\frac{\epsilon - s \delta}{\epsilon + s \delta}} \left[J_{l+1}(k r) + \frac{\left(s - 1\right) l}{k r} J_l (k r)\right]}
	\end{pmatrix}
\end{equation}
\end{widetext}
where  $C_1$ is a normalization constant.
Now for $r>R$ outside the quantum dot,  we  set the  change of variable $\rho = \frac{r}{l_B}$ and express  Eq. \eqref{eqdif} as
\begin{equation}\label{99}
	\left(\frac{\partial^2}{\partial \rho^2}+\frac{1}{\rho} \frac{\partial}{\partial \rho}- \frac{j^2}{\rho^2} -\frac{1}{4} \rho^2-\left(j + s\right)  +\left(k l_B \right)^2\right) \Phi_{A, s}=0
\end{equation}
which can be solved by considering  the following ansatz
\begin{equation}\label{Ans}
\Phi_{A, s}(\rho)=\rho^{\left|j\right|} e^{-\frac{\rho^2}{4}} \xi\left(\rho^2\right). 
\end{equation}
After setting  $\zeta=\frac{\rho^2}{2}$ and substituting Eq. \eqref{Ans} into Eq. \eqref{99},  we arrive at the confluent hypergeometric ordinary differential equation
\begin{equation}
\left[\zeta \frac{\partial^2}{\partial \zeta^2}+(\beta-\zeta) \frac{\partial}{\partial \zeta}-\gamma\right] \xi=0
\end{equation}
where we have defined the parameters
\begin{equation}
	\beta=1+\left|j\right|, \quad \gamma= \frac{\left|j\right|+j+1+s-\left(k l_B\right)^2}{2}.
\end{equation} 
After a straightforward calculation, we obtain the following solutions for the spinors $\Psi_{out, s}(r, \theta) $ outside of the quantum dot
\begin{widetext}	
\begin{equation}
	\Psi_{out, s}=  \frac{C_2 r^{|j|}}{l_B^{|j|}} e^{-\frac{r^2}{4 l_B^2}} e^{i\left(j+\alpha - \phi_i\right) \theta}\begin{pmatrix}
		{ }_1 F_1\left(\gamma, \beta, \frac{r^2}{2 l_B^2}\right) \\
		 \frac{i e^{i s \theta}}{\left( \epsilon + s \delta\right) r  } \left[\left( - |j|+ (s+1) \frac{r^2}{2 l_B^2} +sj\right) { }_1 F_1\left(\gamma, \beta, \frac{r^2}{2 l_B^2}\right) -\frac{r^2}{l_B^2} \frac{\gamma}{\beta} 	{ }_1 F_1\left(\gamma +1, \beta +1, \frac{r^2}{2 l_B^2}\right) \right]
	\end{pmatrix}.
\end{equation}
\end{widetext}
Now to establish the eigenvalues, we employ the continuity of the eigenspinor  at the  quantum point boundary $r=R$. This process yields
\begin{widetext}
\begin{eqnarray}\label{eq17}\nonumber
	&&  \frac{J_l(k R)}{\left( \epsilon + s \delta\right) R  } \left[\left( - |j|+ (s+1) \frac{R^2}{2 l_B^2} +sj\right) { }_1 F_1\left(\gamma, \beta, \frac{R^2}{2 l_B^2}\right) -\frac{R^2}{l_B^2} \frac{\gamma}{\beta} 	{ }_1 F_1\left(\gamma +1, \beta +1, \frac{R^2}{2 l_B^2}\right) \right] \\
	&& -\ { }_1 F_1\left(\gamma, \beta, \frac{R^2}{2 l_B^2}\right)  \sqrt{\frac{\epsilon - s \delta}{\epsilon + s \delta}} \left[J_{l+1}(k R) + \frac{\left(s - 1\right) l}{k R} J_l (k R)\right] =0.
\end{eqnarray}
\end{widetext}
It is noteworthy that solving Eq. \eqref{eq17} analytically is a challenging task. Therefore, resorting to a numerical solution becomes necessary. In this regard, we will present various numerical analyses of Eq. \eqref{eq17}, exploring its fundamental characteristics to extract more pertinent information about the current system.

\section{RESULTS AND DISCUSSION} \label{results}

\begin{figure*}[ht]
	\centering
	\includegraphics[scale=0.34]{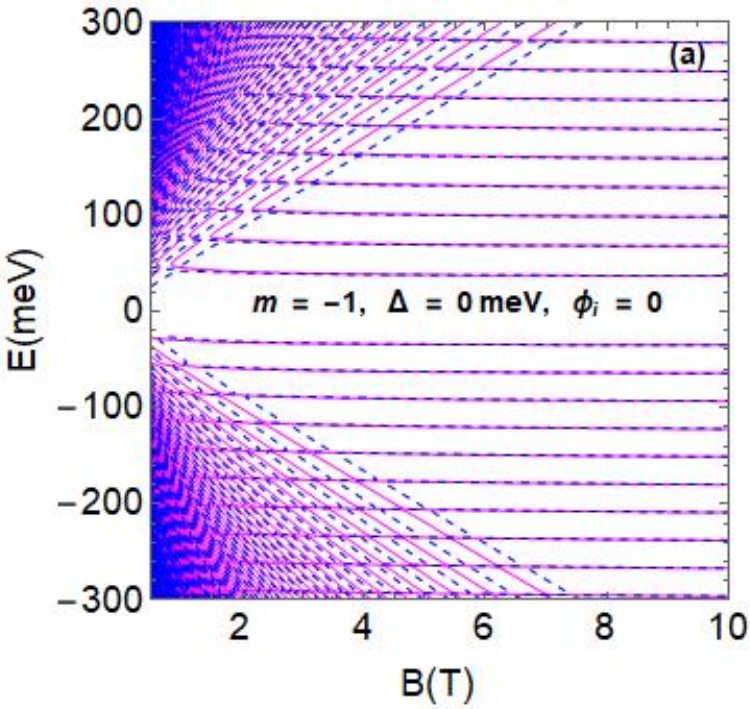}
	\includegraphics[scale=0.34]{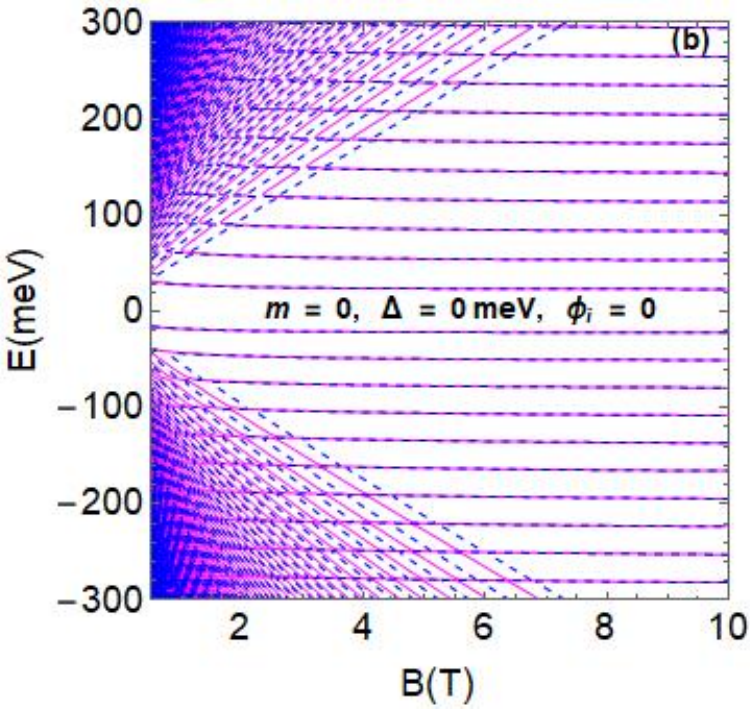}
	\includegraphics[scale=0.34]{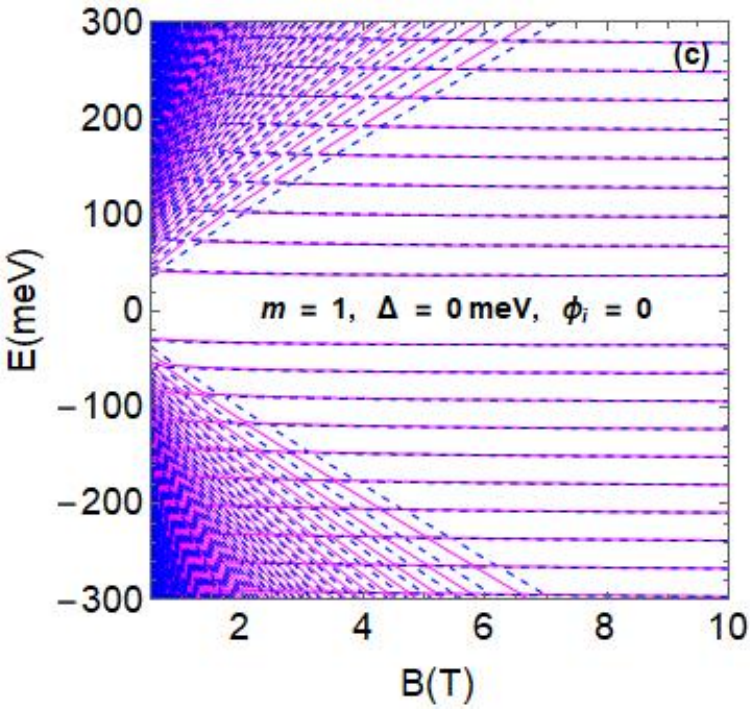}	
	\includegraphics[scale=0.34]{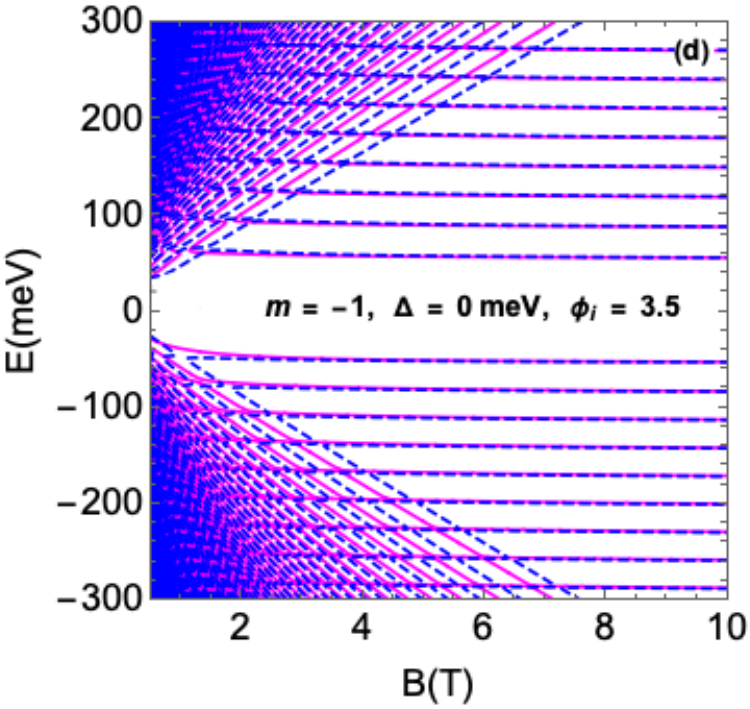}
	\includegraphics[scale=0.34]{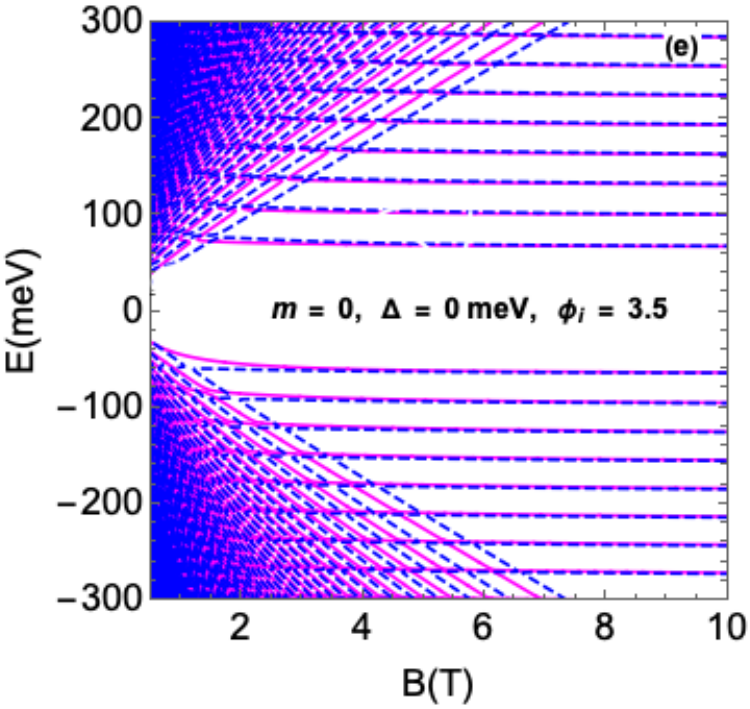}
	\includegraphics[scale=0.34]{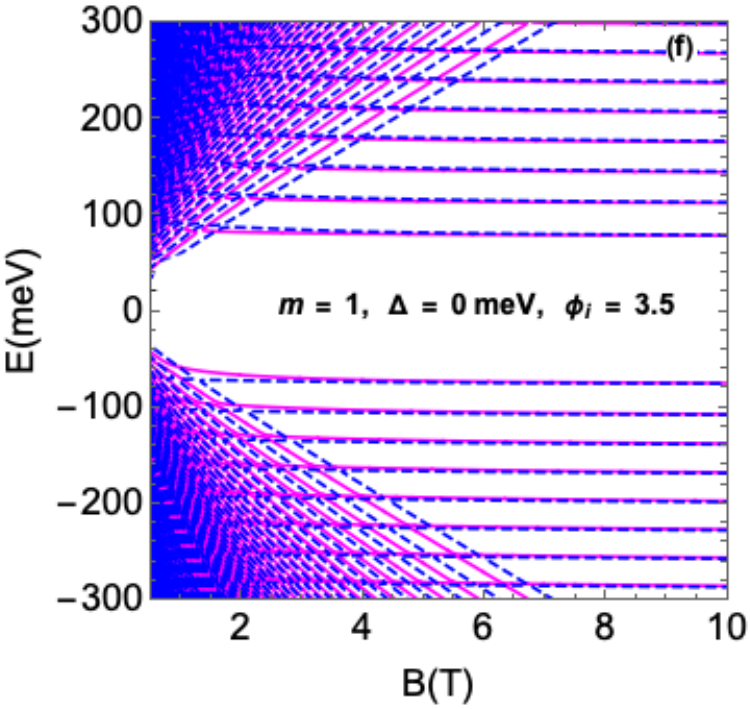}	\includegraphics[scale=0.34]{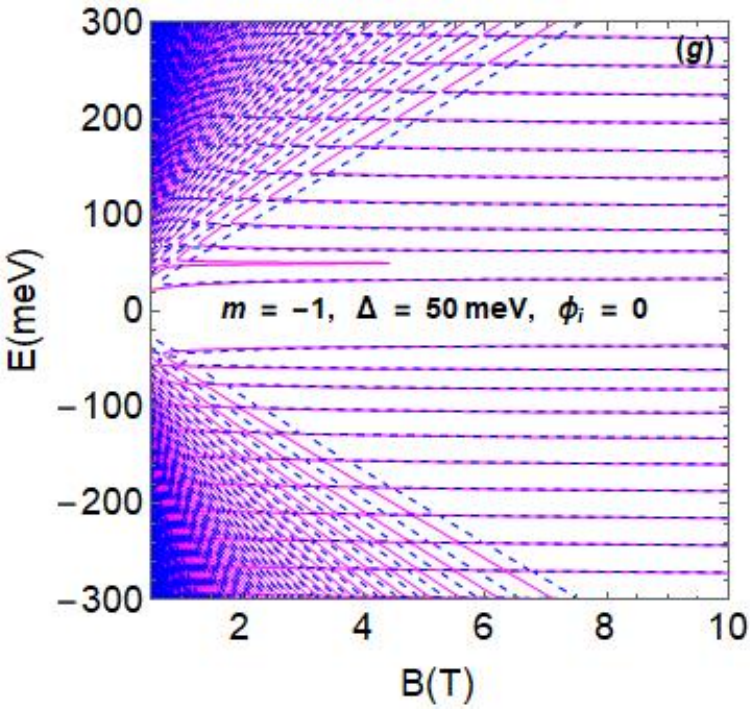}
	\includegraphics[scale=0.34]{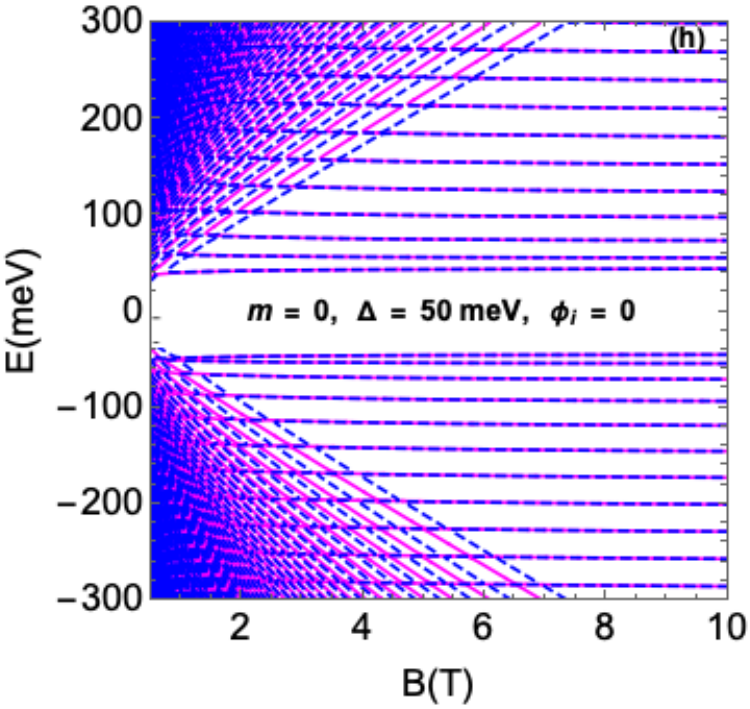}
	\includegraphics[scale=0.34]{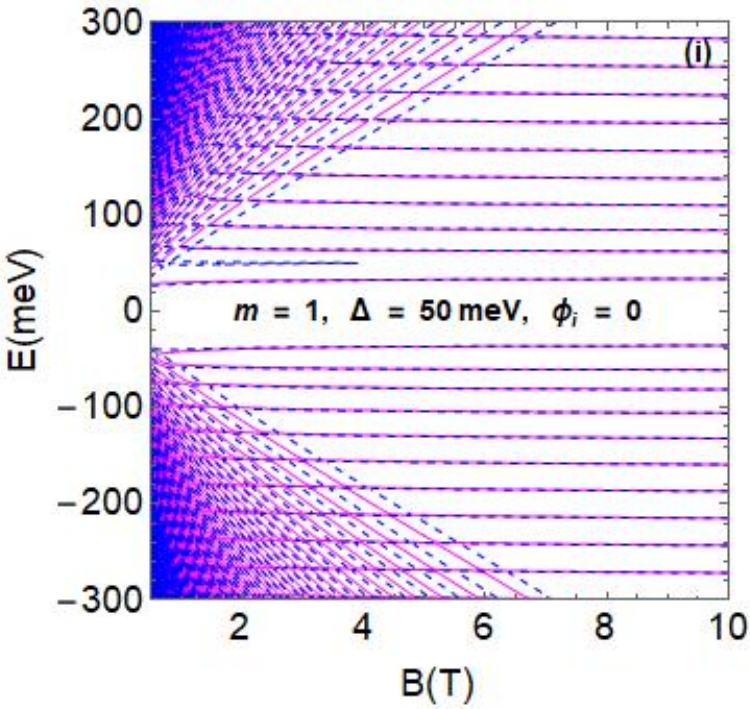}
	\includegraphics[scale=0.34]{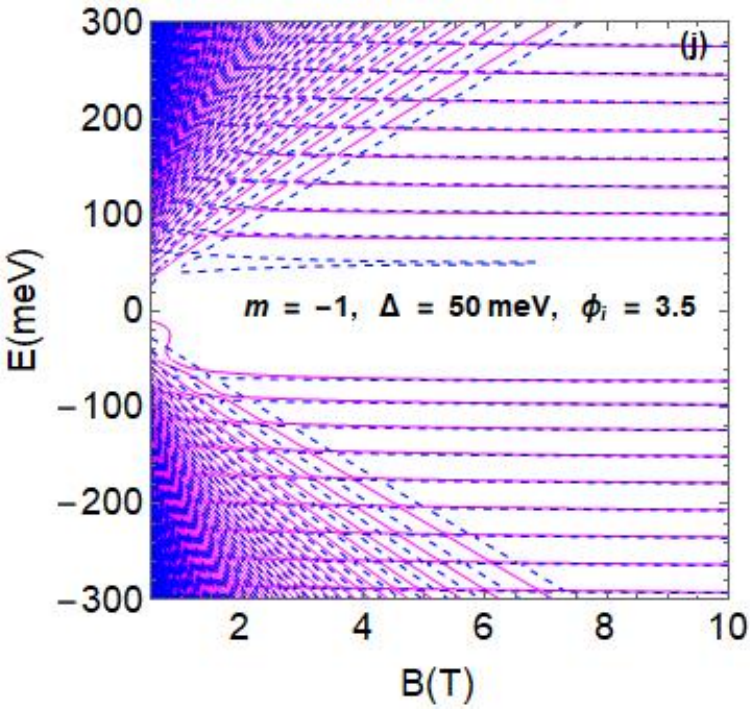}
	\includegraphics[scale=0.34]{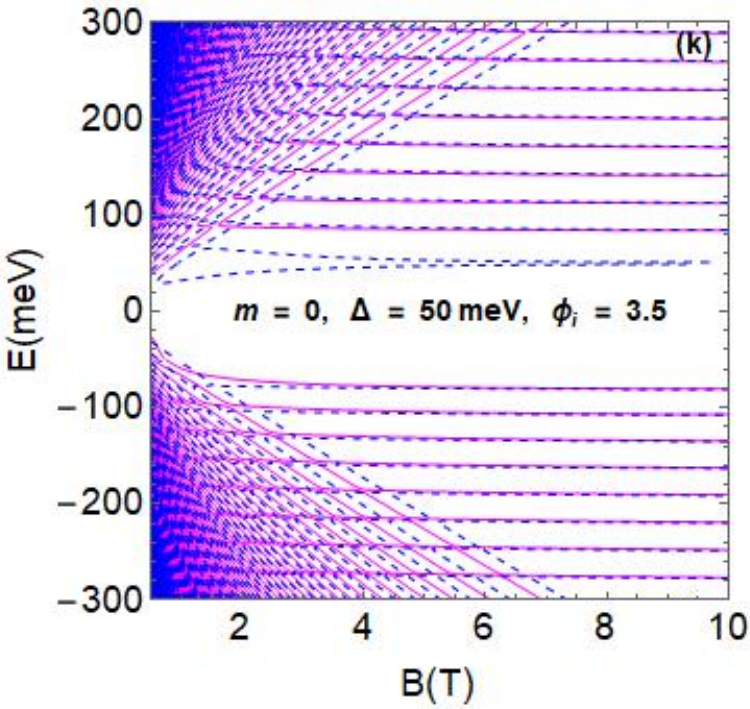}
	\includegraphics[scale=0.34]{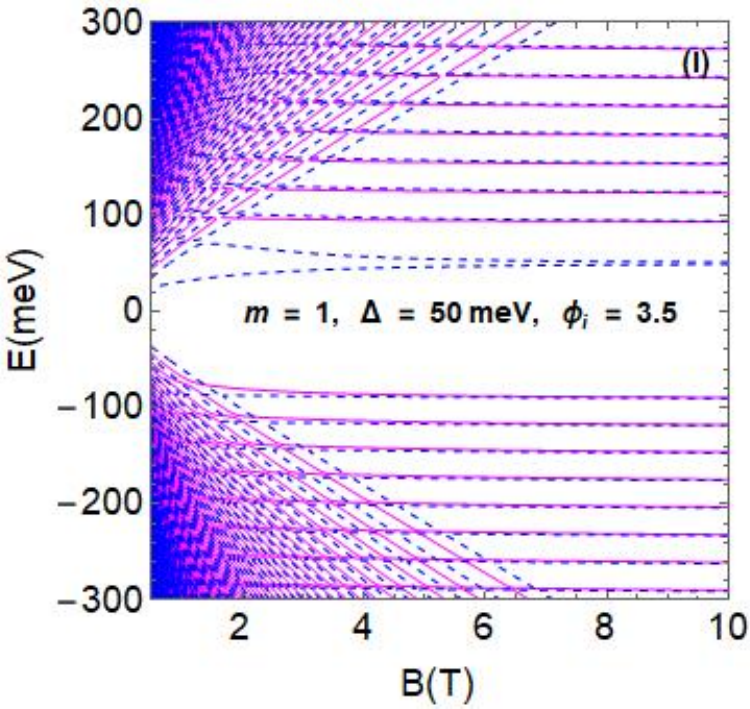}
	\caption{(color online) Energy levels $E$ as a function of the magnetic field $B$ for $R=70$ nm with $m=-1$ (a,d,g,j), $m=0$ (b,e,h,k), $m=1$ (c,f,i,l) by changing the values of energy gap $\Delta$ and Aharonov-Bohm flux $\phi_i$ such that (a,b,c) for ($\Delta = 0$ meV, $\phi_i = 0$), (d,e,f) for ($\Delta = 0$ meV, $\phi_i = 3.5$), (g,h,i) for ($\Delta = 50$ meV, $\phi_i = 0$), and (j,k,l) for ($\Delta = 50$ meV, $\phi_i = 3.5$). Dashed blue curves for the valley $K$ ($s=1$) and magenta curves for the valley $K'$ ($s=-1$).}
	\label{fig2}
\end{figure*}

Fig. \ref{fig2} illustrates the energy levels as a function of the magnetic field $B$ for $R = 70$ nm, while varying the values of Aharonov-Bohm flux $\phi_i$ and energy gap $\Delta$. Adjacent panels result from changing the value of  parameters are listed as follows:  Specifically, the panels correspond to: (a) $m = -1$, (b) $m = 0$, (c) $m = 1$ for ($\Delta = 0$ meV, $\phi_i = 0$); (d) $m = -1$, (e) $m = 0$, (f) $m = 1$ for ($\Delta = 0$ meV, $\phi_i = 3.5$); (g) $m = -1$, (h) $m = 0$, (i) $m = 1$ for ($\Delta = 50$ meV, $\phi_i = 0$); and (j) $m = -1$, (k) $m = 0$, (l) $m = 1$ for ($\Delta = 50$ meV, $\phi_i = 3.5$).
For $\phi_i = 0$ and $\Delta = 0$, when a weak magnetic field tends towards zero ($B \to 0$), we observe that the energy levels exhibit a continuous energy band as depicted in panels (a,b,c). This results in multiple degenerate states for all angular momentum \(m\) and for both valleys \(K\) and \(K^\prime\).
For a very strong magnetic field, we note that the degeneracy is lifted while maintaining symmetry, allowing us to express \textbf{$E(m, s) = E(m, -s)$}. This finding aligns with previous research results \cite{mirzakhani2016energy, belokda2022energy}.
Examining panels (d,e,f) of Fig. \ref{fig2}, the dependence of the energy spectrum on the magnetic field is evident with $R = 70$ nm and angular momenta $m = -1$ (d), $m = 0$ (e), and $m = 1$ (f) for the two valleys: $K$ illustrated by the blue dashed curves, and $K^\prime$ shown by the magenta solid curves. These outcomes were obtained without introducing gap energy in the presence of magnetic flux within the graphene quantum dot. A subtle separation of the energy spectrum band, initially continuous at low field strengths, is observed, along with a notable opening of an energy gap between the conduction and valence bands.
Analyzing panels (g,h,i) of Fig. \ref{fig2}, it becomes evident that a new energy level emerges at non-zero angular momentum ($m \neq 0$) when the magnetic field strength is less than 4 ($B \leq 4$), leading to perturbations in the energy spectrum, as seen in panels (g,i). When magnetic flux and gap energy are combined, the resulting band gap becomes larger, see  panels (j,k,l). Notably, a new energy level emerges within the band gap for all values of angular momentum $m$. The appearance of this new energy level is particularly conspicuous when $m=1$, as observed in panel (l) of Fig. \ref{fig2}. Importantly, this emergence is only observable when $s$ is equal to one. The degeneracy of energy levels is lost under the influence of a strong magnetic field applied perpendicularly to the graphene sheet. As the magnetic field  increases, the energy spectrum of charge carriers increasingly merges into the quantum dots. At the center of the quantum dot, carriers become highly localized.

\begin{figure*}[ht]
	\centering
	\includegraphics[scale=0.34]{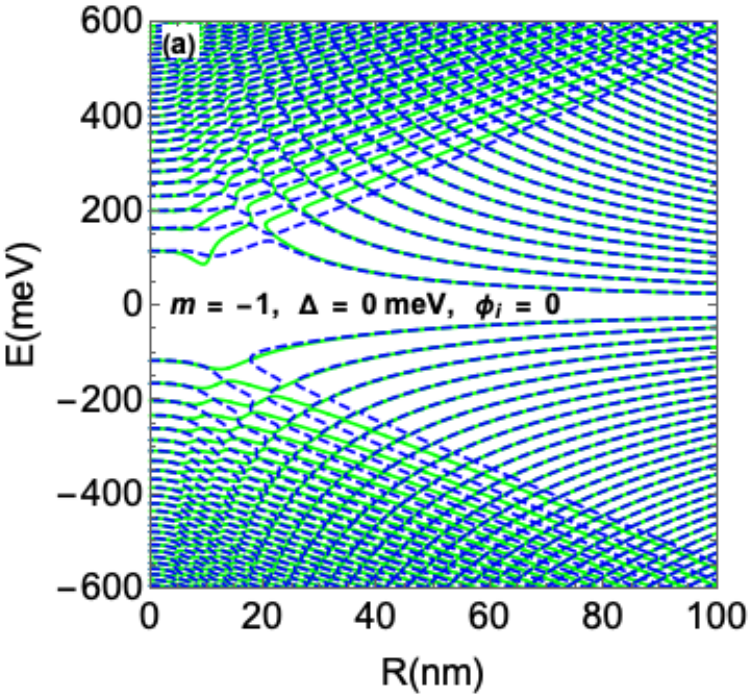}
	\includegraphics[scale=0.34]{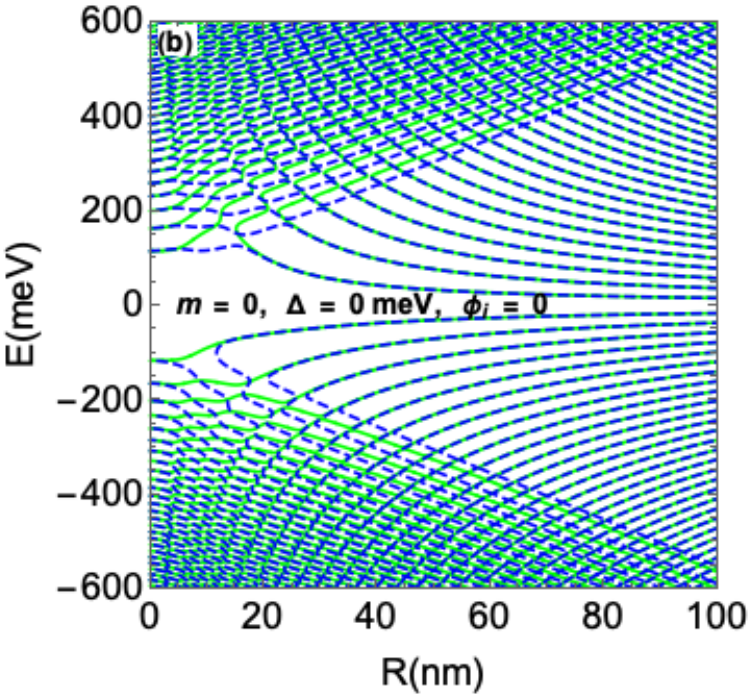}
	\includegraphics[scale=0.34]{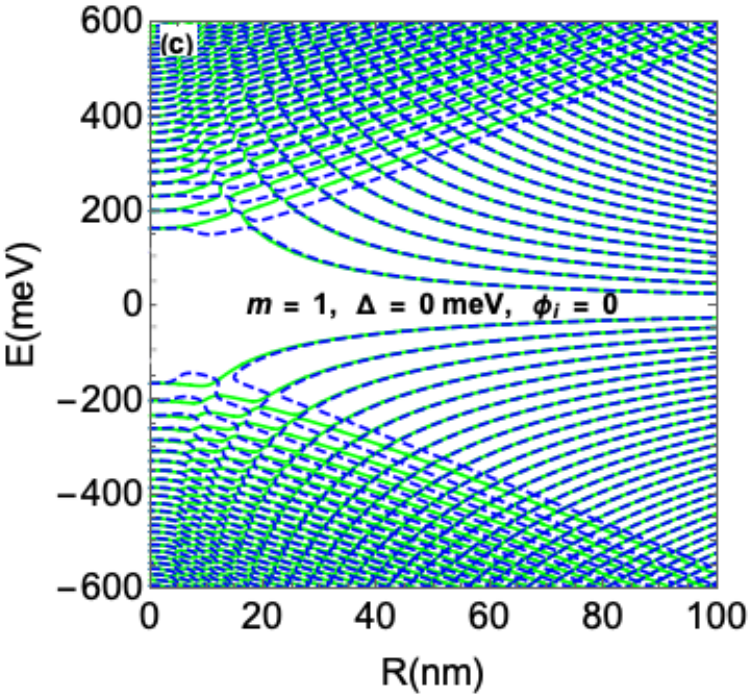}	\includegraphics[scale=0.34]{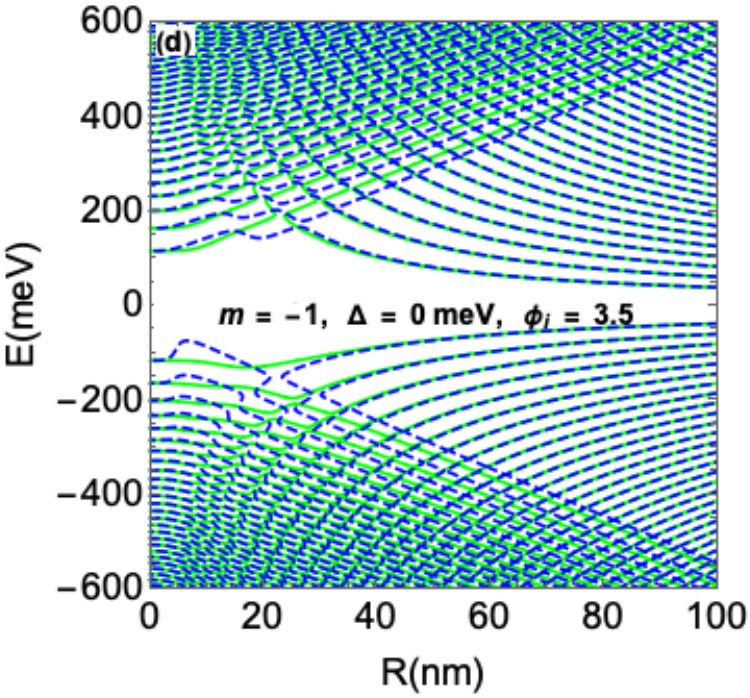}
	\includegraphics[scale=0.34]{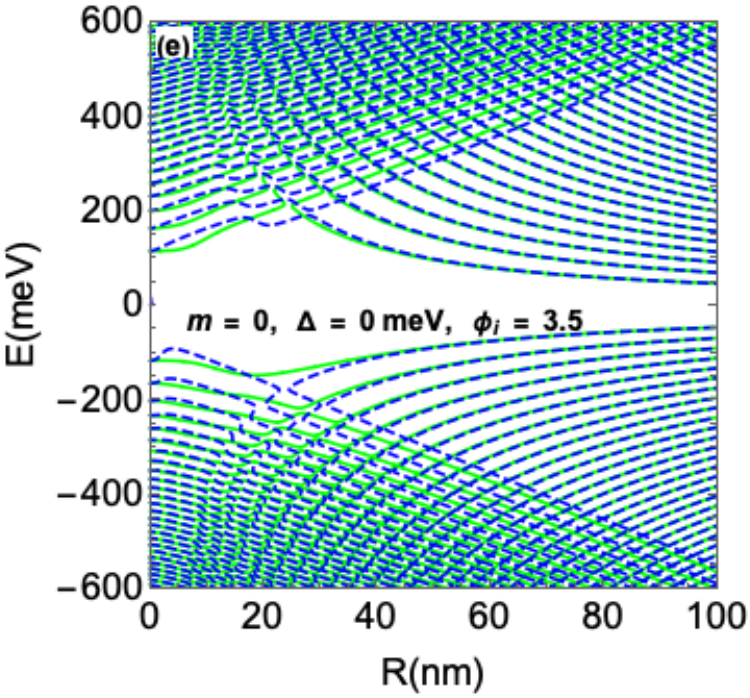}
	\includegraphics[scale=0.34]{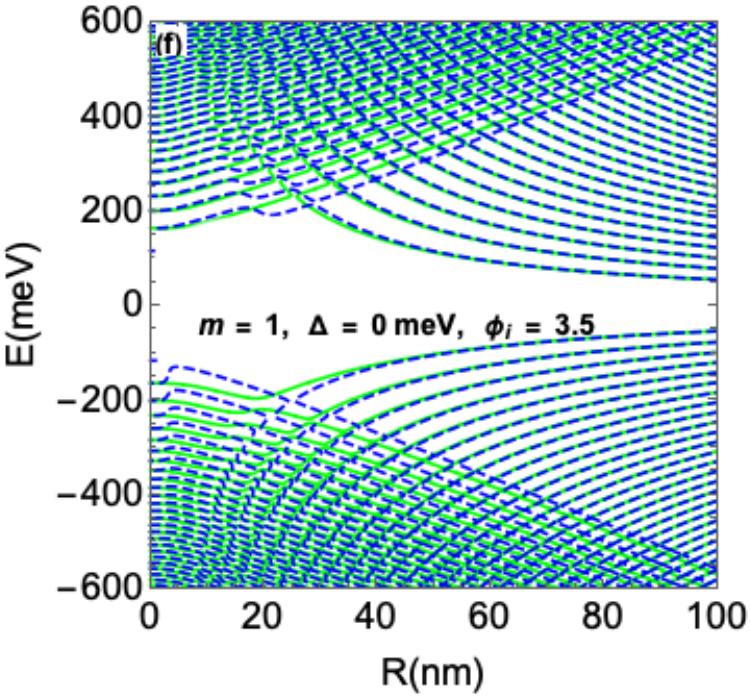}	\includegraphics[scale=0.34]{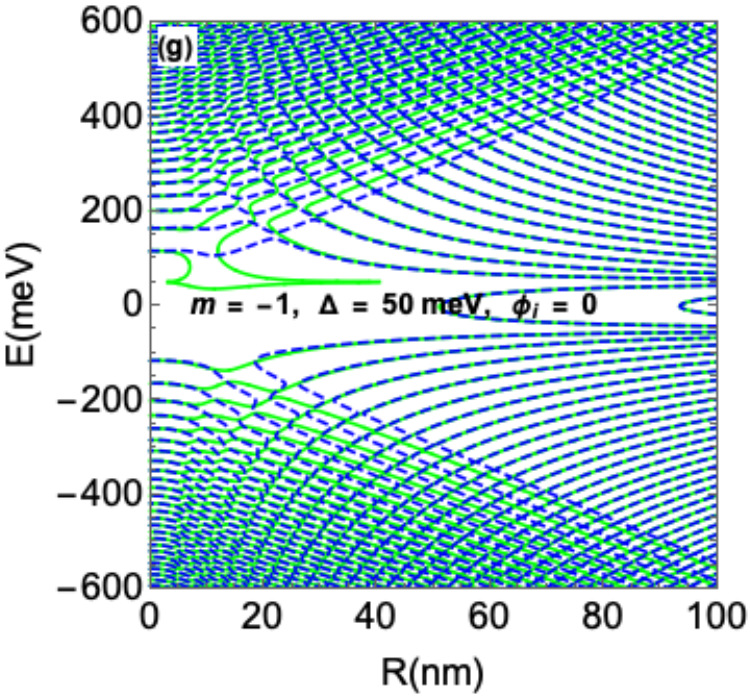}
	\includegraphics[scale=0.34]{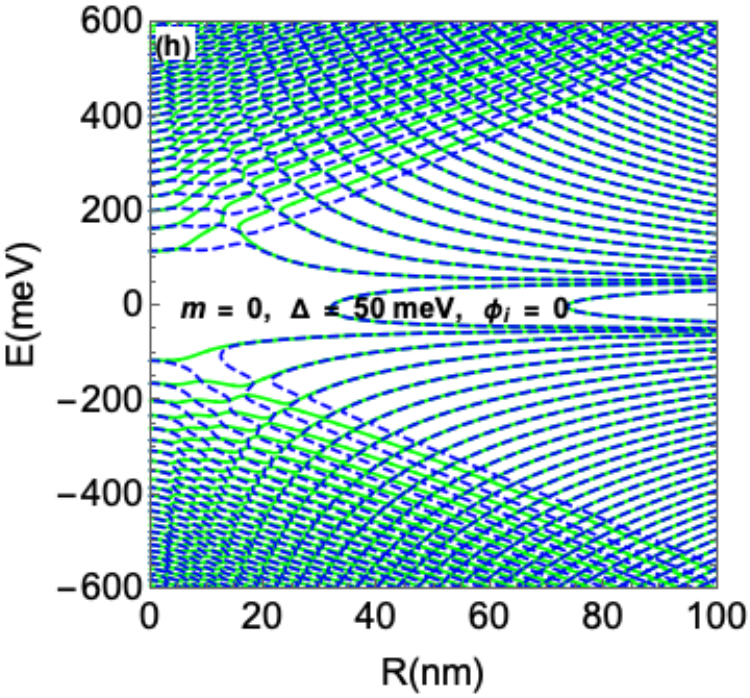}
	\includegraphics[scale=0.34]{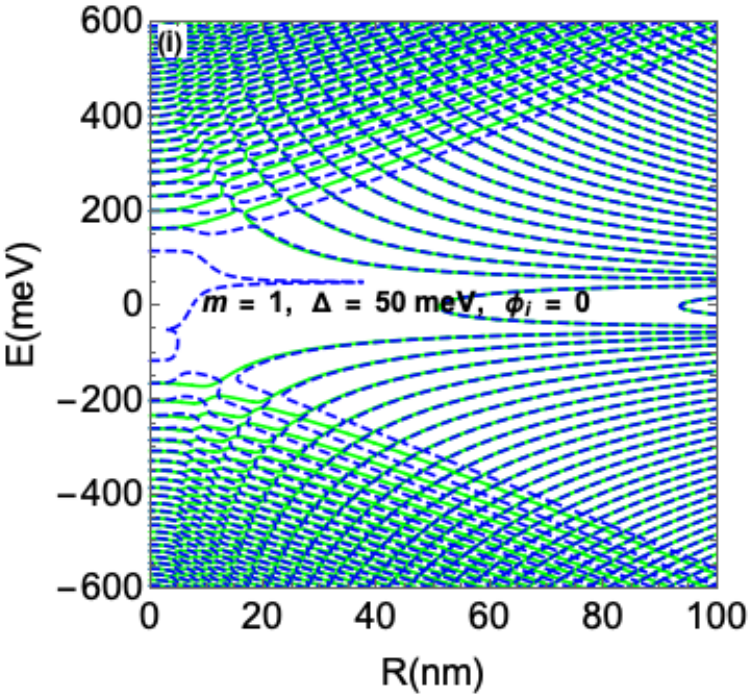}
	\includegraphics[scale=0.34]{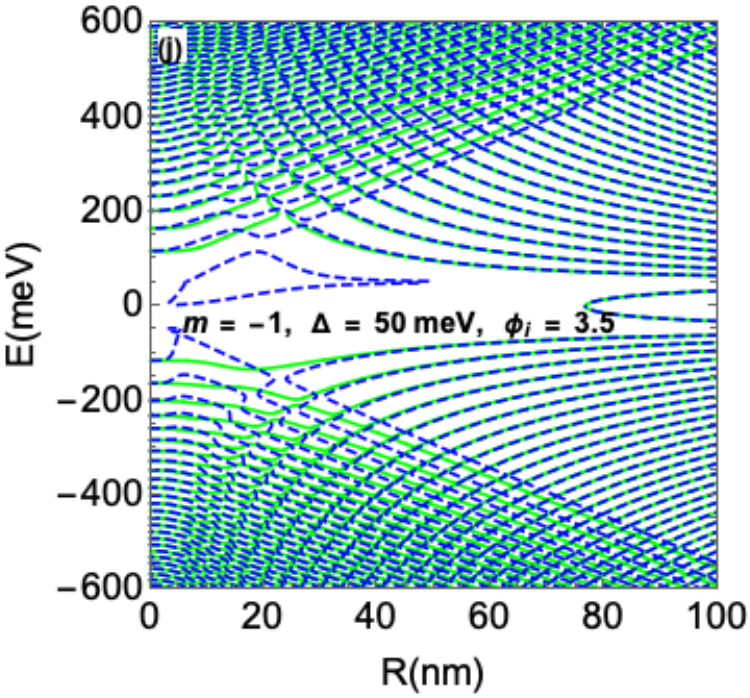}
	\includegraphics[scale=0.34]{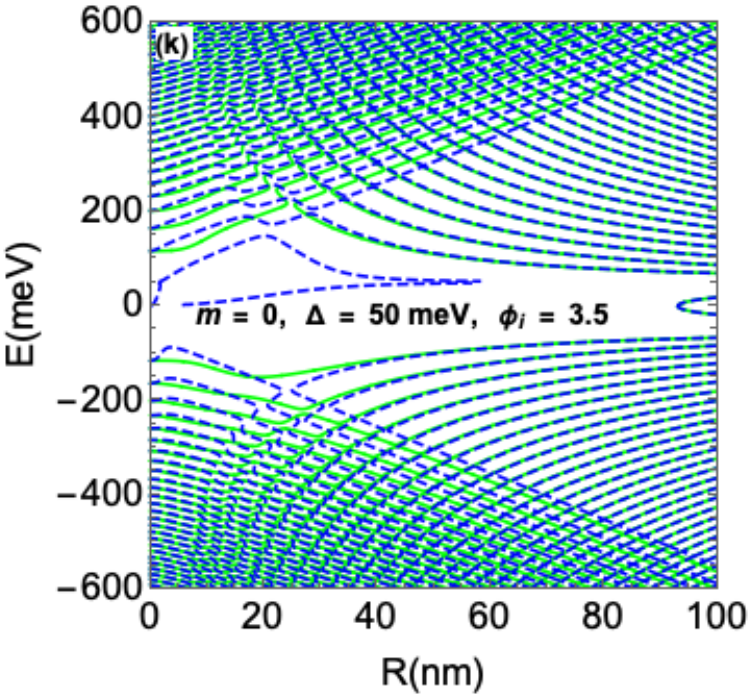}
	\includegraphics[scale=0.34]{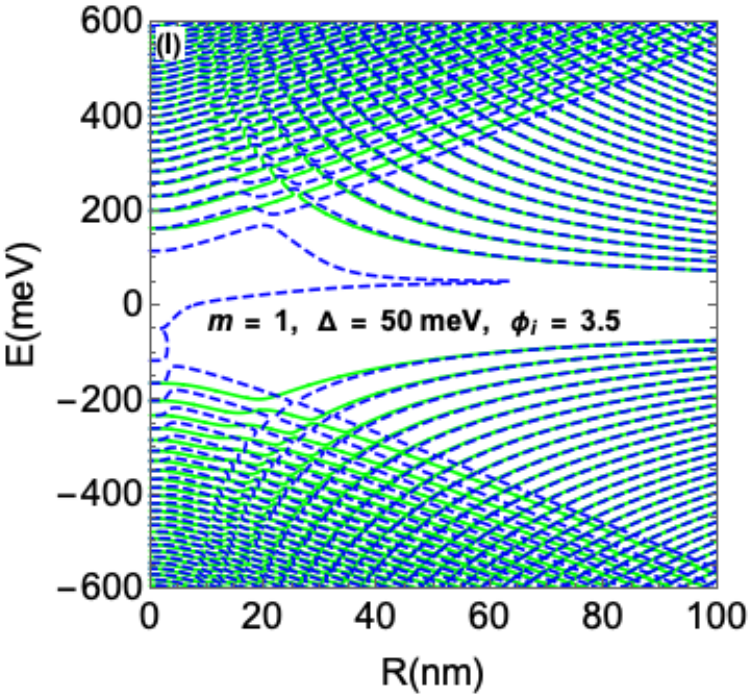}
	\caption{(color online) Energy levels $E$ as a function of the quantum dot radius $R$ for $B = 10$ T with $m=-1$ (a,d,g,j), $m=0$ (b,e,h, k), $m=1$ (c,f,i,l) by changing the values of energy gap $\Delta$ and Aharonov-Bohm flux $\phi_i$: (a,b,c) for ($\Delta = 0$ meV, $\phi_i = 0$), (d,e,f) for ($\Delta = 0$ meV, $\phi_i = 3.5$), (g, h, i) for ($\Delta = 50$ meV, $\phi_i = 0$), and (j,k,l) for ($\Delta = 50$ meV, $\phi_i = 3.5$). Dashed blue curves for the valley $K$ ($s=1$) and green curves for the valley $K'$ ($s=-1$).}
	\label{fig3}
\end{figure*}

\begin{figure}[ht]
	\centering
	\includegraphics[scale=0.33]{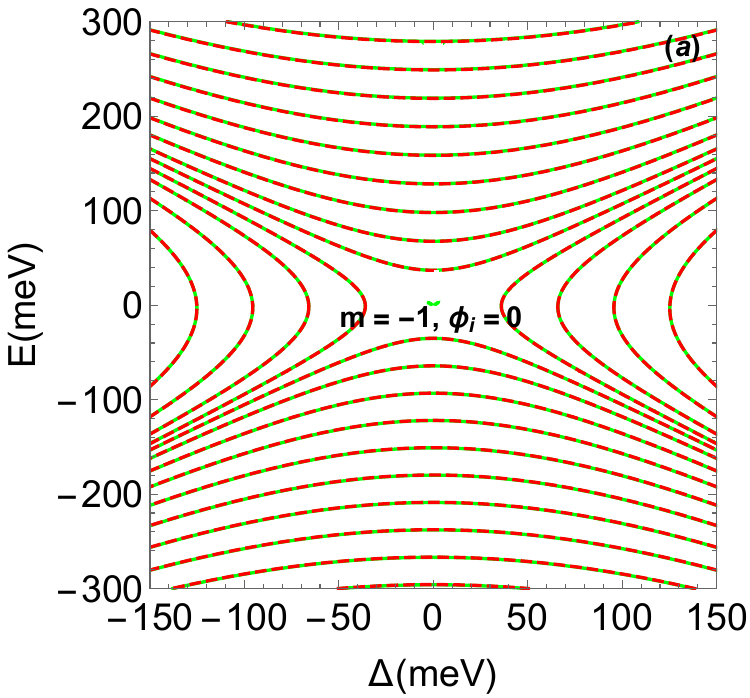}
	\includegraphics[scale=0.33]{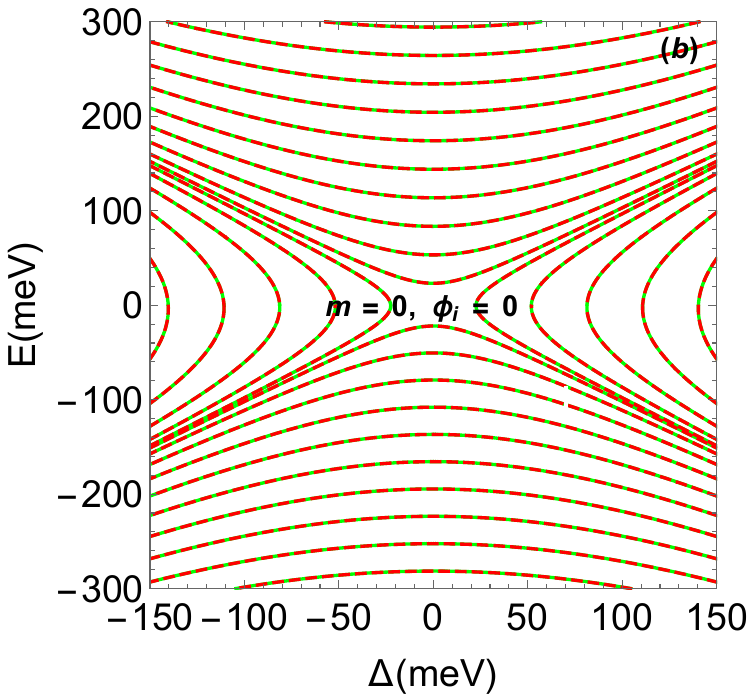}\\
	\includegraphics[scale=0.33]{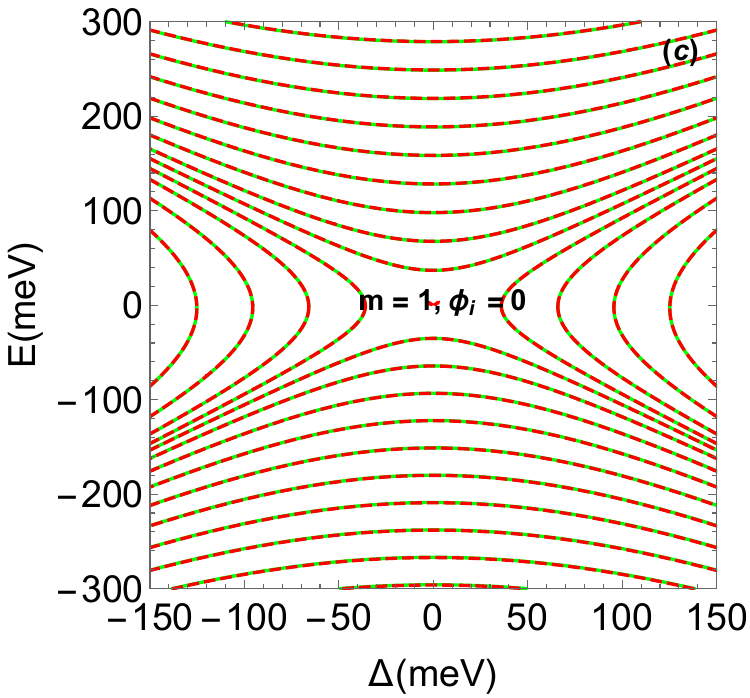}	\includegraphics[scale=0.33]{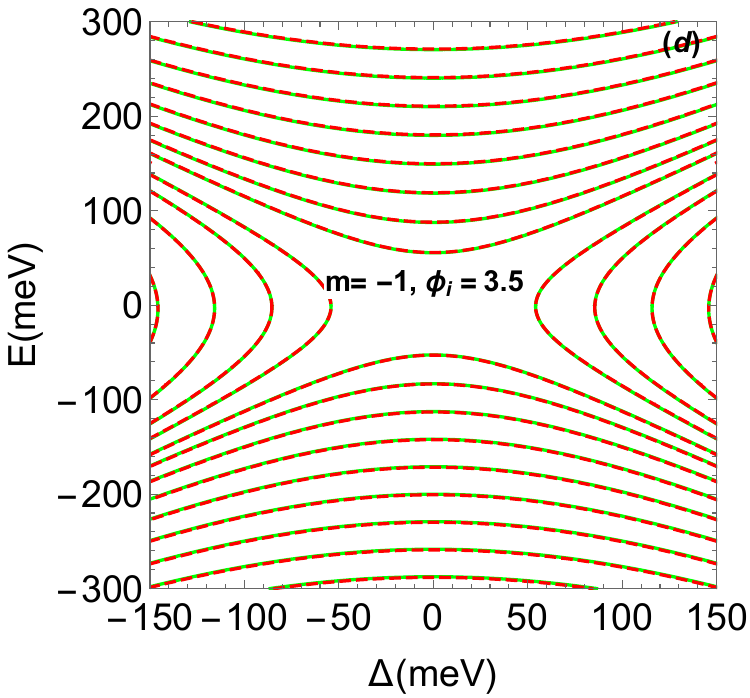}\\
	\includegraphics[scale=0.33]{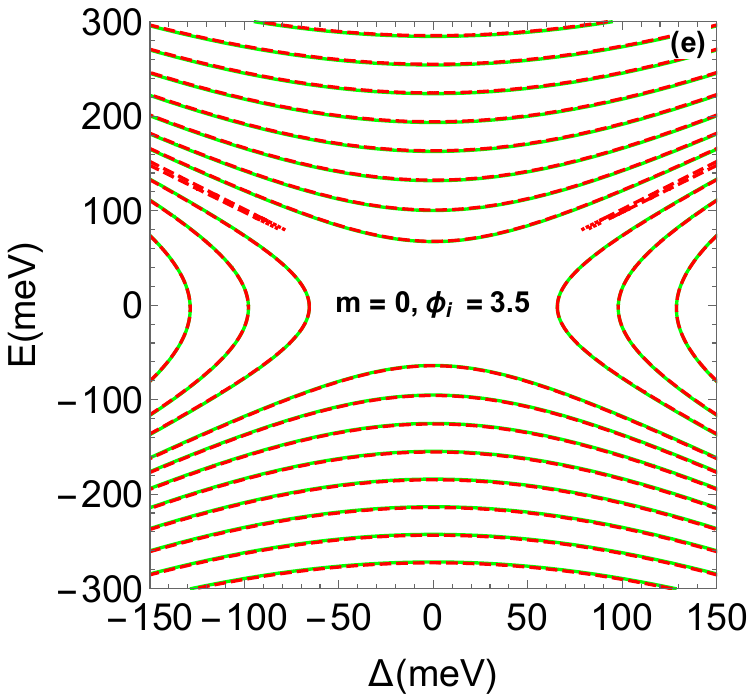}
	\includegraphics[scale=0.33]{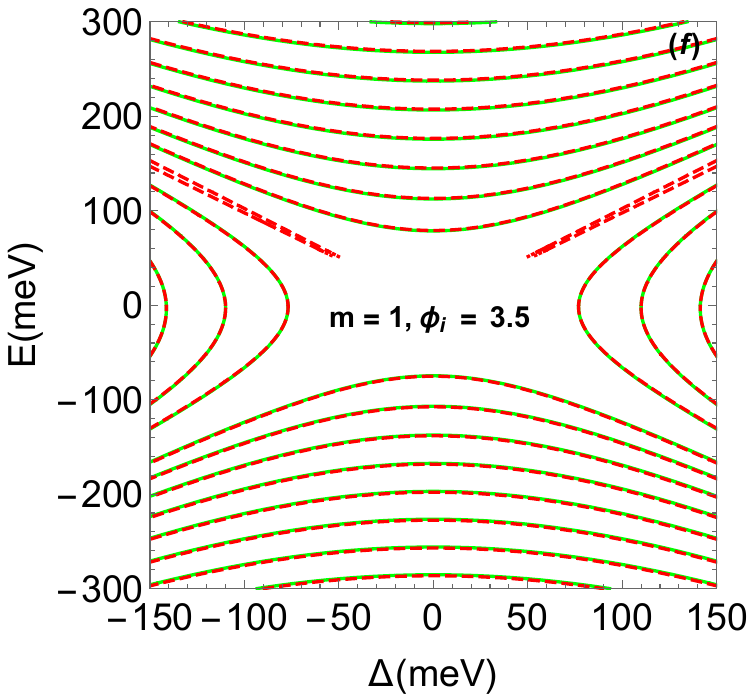}
	\caption{(color online) Energy levels $E$ as a function of the energy gap $\Delta$ for $B = 10$ T and $R=70$ nm with $m=-1$ (a,d), $m=0$ (b,e), $m=1$ (c,f) by varying the value of the Aharonov-Bohm flux $\phi_i$: (a,b,c) for $\phi_i = 0$ and (d,e,f) for $\phi_i = 3.5$. Dashed red curves for the valley $K$ ($s=1$) and green curves for the valley $K'$ ($s=-1$).}
	\label{fig4}
\end{figure}

Fig. \ref{fig3} shows the relationship between graphene quantum dot energy levels and radius $R$ for magnetic field strength $B= 10 T$.There are three different values (-1, 0 and 1) for the angular momentum $m$. This dependence is based on changes in two external parameters: the gap energy $\Delta$ and the magnetic flux value $\phi_i$. In a quantum dot without applied magnetic flux or gap energy, see panels (a,b,c) of Fig. \ref{fig3}), an energy gap emerges between the electron and hole energy levels. As the radius of the quantum dot increases, this gap closes. We observe that when the value of $R$ is very small, the energy levels for the two valleys $K$ and $K^\prime$ become almost equal. This leads to the symmetry of the system, where $E(m, s)= -E(m, -s)$. In other cases, the system shows an asymmetry where $E(m, s) = E(m, -s)$. After the inclusion of the magnetic flux term in our system, we observe a significant increase in the band gap between the conduction and valence bands, leading to improved control of the electron flow, as shown by the results in panels (d,e,f) of Fig. \ref{fig3}. As shown in Fig. \ref{fig3} (g,h,i), adding a gap energy value of $\Delta = 50$ meV creates new energy levels between the valence and conductance bands for all values of angular momentum $m = 0, \pm 1$. The new energy levels create an asymmetry in the energy spectrum, consistent with \cite{mirzakhani2016energy, belokda2022energy}. Our system is affected by two external parameters: the gap energy and the magnetic flux. These parameters are non-zero inside the graphene quantum dot. In  panels (j,k,l) of Fig. \ref{fig3}, the results of the numerical simulation are presented when both terms are included simultaneously. These results indicate that the magnetic flux causes an increase in the opening of the band gap and a decrease in the energy levels between the conduction band and the valence band. In addition, the symmetry of the energy spectrum is disturbed.

 \begin{figure}[ht]
	\centering
	\includegraphics[scale=0.33]{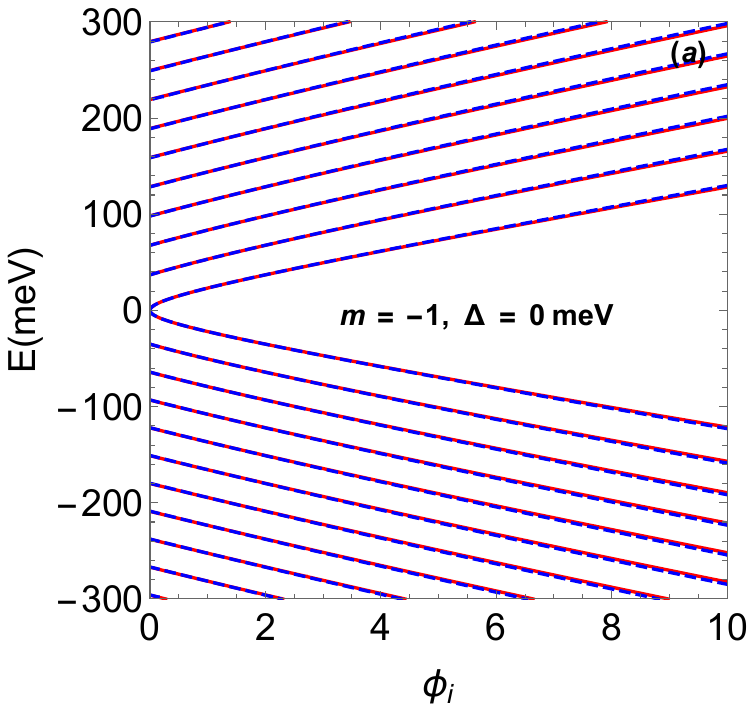}
	\includegraphics[scale=0.33]{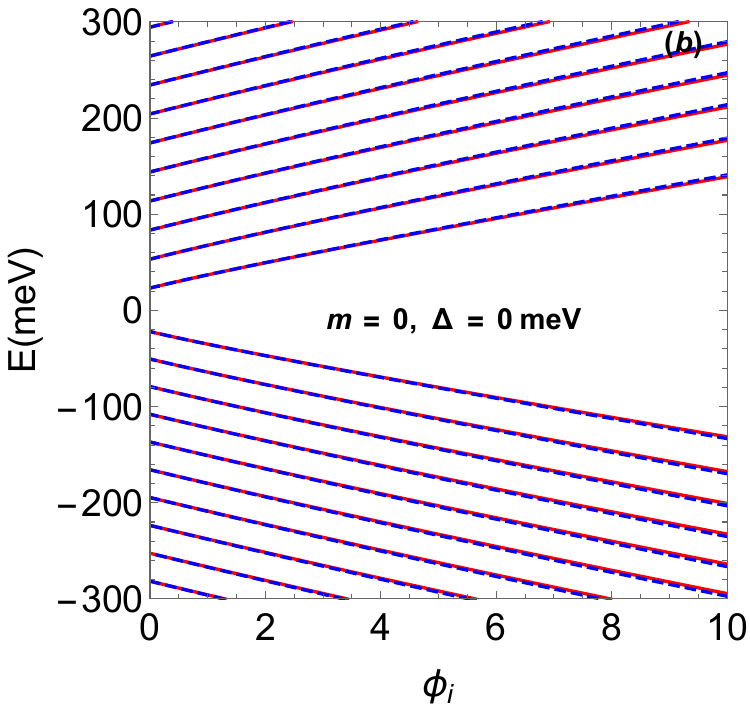}\\
	\includegraphics[scale=0.33]{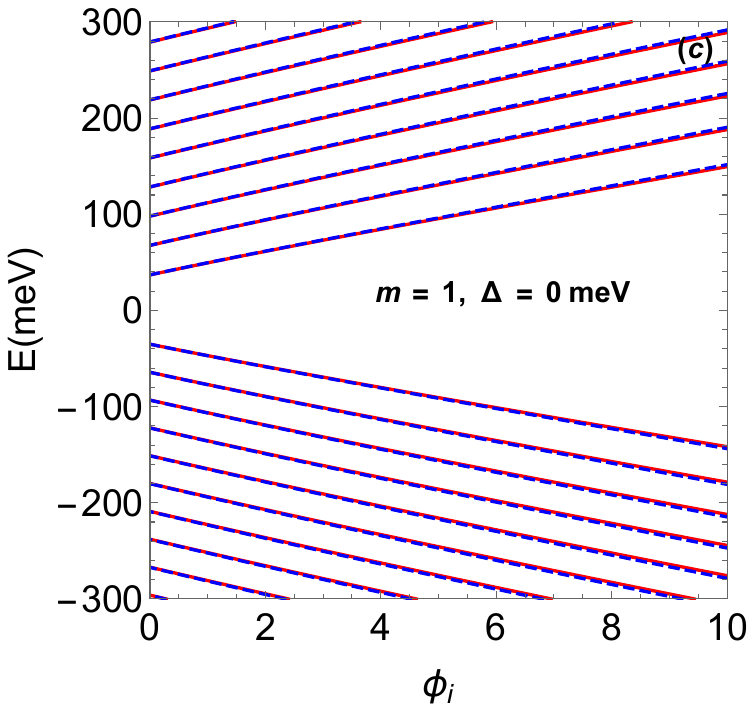}	\includegraphics[scale=0.33]{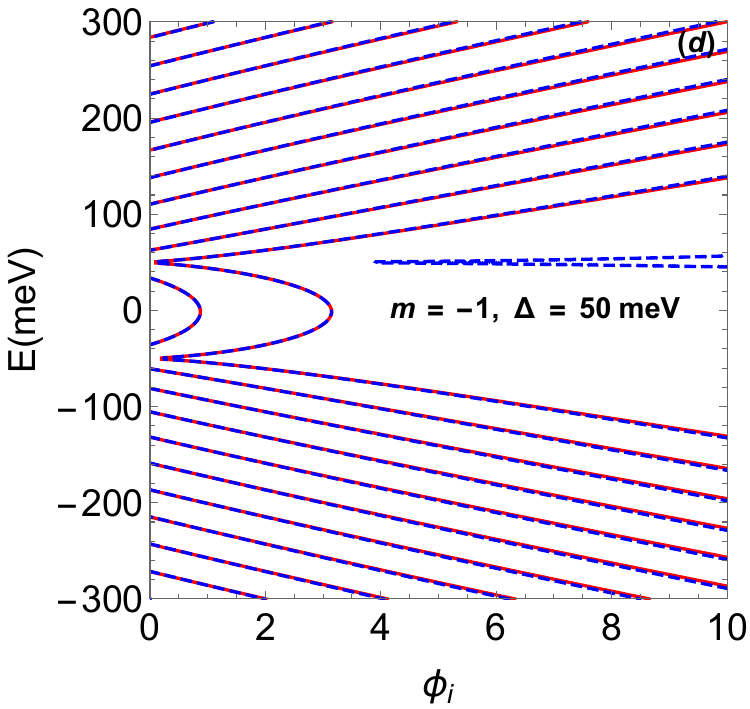}\\
	\includegraphics[scale=0.33]{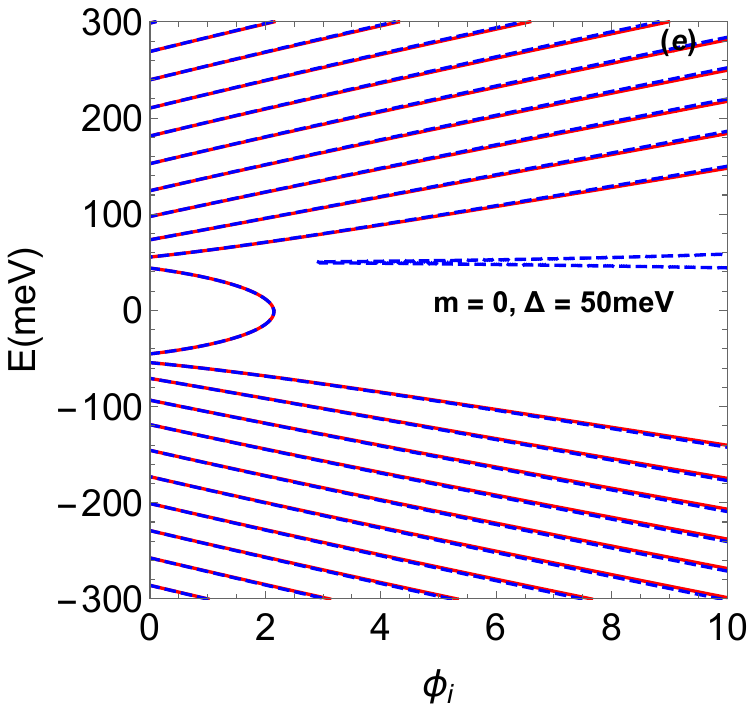}
	\includegraphics[scale=0.33]{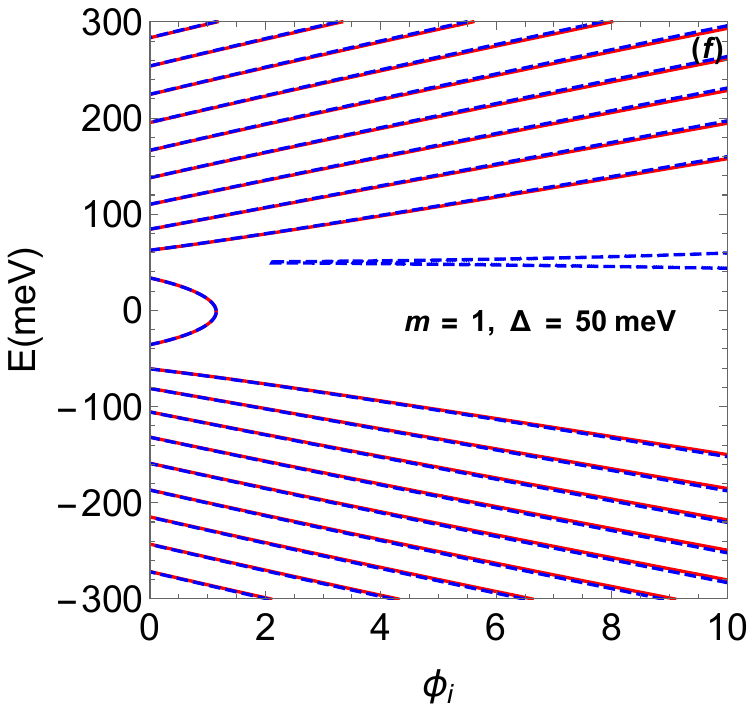}
	\caption{(color online) Energy levels $E$ as a function of the Aharonov-Bohm flux  $\phi_i$ for  $B = 10$ T and $R=70$ nm with $m=-1$ (a,d), $m=0$ (b,e), $m=1$ (c,f) by varying the energy gap $\Delta$: (a,b,c) for $\Delta = 0$ meV and (d,e,f) for $\Delta = 50$ meV. Dashed blue curves for the valley $K$ ($s=1$) and red curves for the valley $K'$ ($s=-1$).}
	\label{fig5}
\end{figure}

Fig. \ref{fig4} depicts the evolution of the energy spectrum concerning the gap energy $\Delta$ while keeping all other parameters constant and varying the Aharonov-Bohm flux $\phi_i$, with $B = 10$ T and $R = 70$ nm. Here, the angular momentum is fixed at 0, -1, and 1. Considering the two values of the $s$ valley, the green curves represent the $K^\prime$ valley ($s=-1$), while the red dashed curves represent the $K$ valley ($s=1$). When the Aharanov-Bohm flux is absent ($\phi_i = 0$), the energy levels exhibit changes with the gap energy, forming horizontal parabolas, confirming the symmetry $E(m,s) = E(m,-s)$. In the presence of an external magnetic field and a non-canceling Aharonov-Bohm flux within the graphene quantum dot, the energy levels obtained by the continuum model at the edges also exhibit parabolic variation as a function of the gap energy $\Delta$. In this scenario, numerous additional energy states are introduced, disrupting energy spectrum symmetries and widening band gap widths. Notably, the impact of an Aharonov-Bohm flux on the gap in graphene energy spectra is particularly significant. This phenomenon is evident in the continuum model, suggesting that energy gaps are more effectively controlled by both an external field applied outside the graphene quantum dot and the presence of a strong Aharonov-Bohm flux applied to the center of our system.

The next step is to investigate the effect of the Aharonov-Bohm flux $\phi_i$ on the energy spectrum of a graphene quantum dot. Specifically, our focus is on the case of a strong magnetic field with an intensity of 10 T and a quantum dot radius of $R = 70$ nm. We choose three angular momentum values: $m = 0, \pm 1$, with the gap energy $\Delta$ set to $0$ meV, as depicted in Figs. \ref{fig5} (a,b,c). As a result, the energy gaps between the valence and conduction bands are found together with perfect symmetry. The energy spectrum varies parabolically with the magnetic flux. For $\Delta$ is 50 meV as represented in Figs. \ref{fig5} (d,e,f), one can see that the presence of an energy gap destroys the symmetry and allows for the creation of new energy levels.

At this level, let us investigate the scenario in which an external magnetic field is taken into account and an Aharonov-Bohm flux is present within a quantum dot. As the magnetic field strength increases, the degenerate levels associated with each $m$ value undergo splitting, and the quantum dot states amalgamate to give rise to Landau levels. Specifically, in the presence of a highly intense magnetic field, the term $\frac{r^2}{2 l_B^2}$ tends towards zero, allowing us to approximate thehypergeometric function by \cite{olver1964handbook}
\begin{equation}
	{ }_1 F_1\left(\alpha, \beta, \gamma\right)=\frac{\Gamma(\beta)}{\Gamma(\alpha)} e^\gamma \gamma^{\alpha-\beta}\left[1+O\left(\mid \gamma^{-1}\right)\right].
\end{equation}
Utilizing the Gamma function to express binomial coefficients and applying the relation $\Gamma(y+1) = y \Gamma(y)$, Eq. \eqref{eq17} can be simplified to
\begin{equation} \label{19}
	E_{mn} = \pm \frac{\hbar v_F}{l_B} \sqrt{2 n^2 + 2 \delta^2 + |j| + j +1 -s}
\end{equation}
where the quantum $n = 1,2,3, \cdots  $ are indexing the Landau levels, and $j = m - \alpha + \phi_i$. 

\begin{figure}[ht]
	\centering
	\includegraphics[scale=0.335]{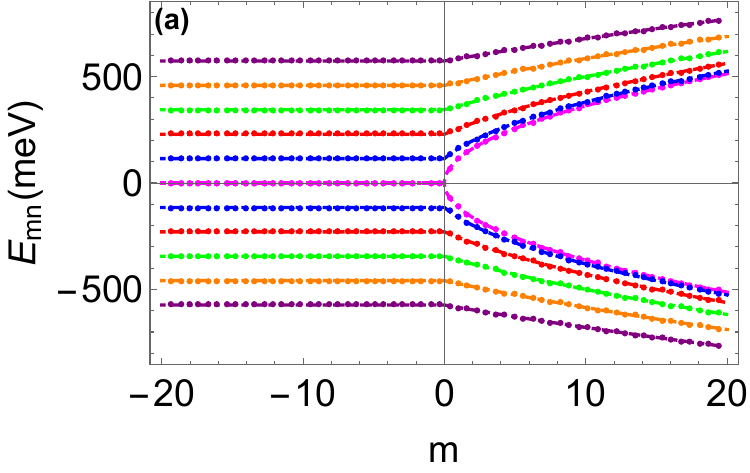}
	\includegraphics[scale=0.335]{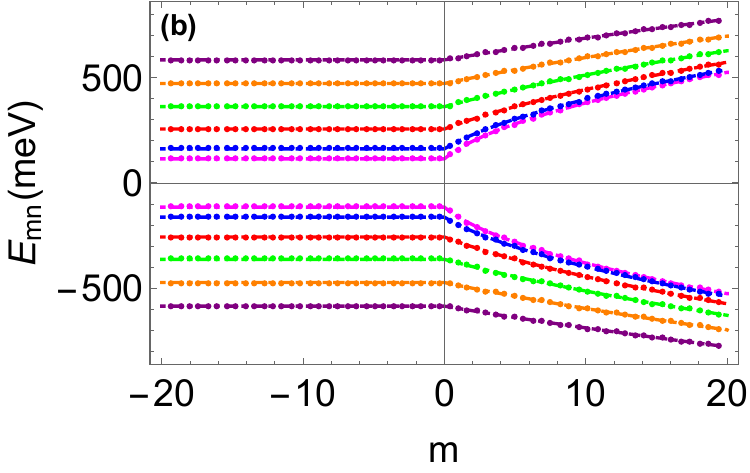}\\	\includegraphics[scale=0.335]{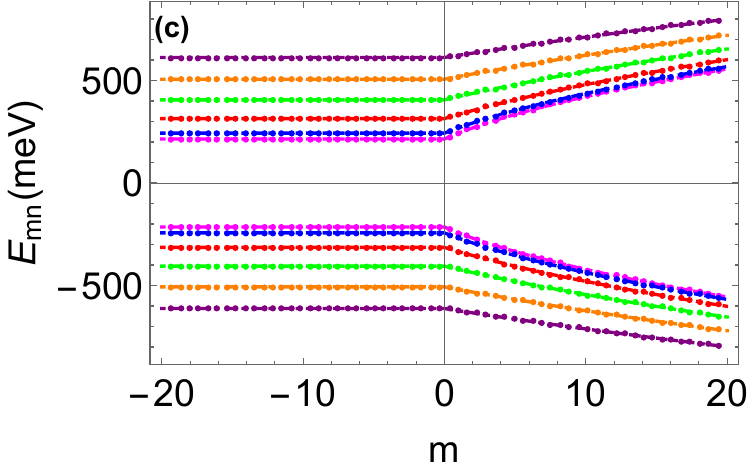}
	\includegraphics[scale=0.335]{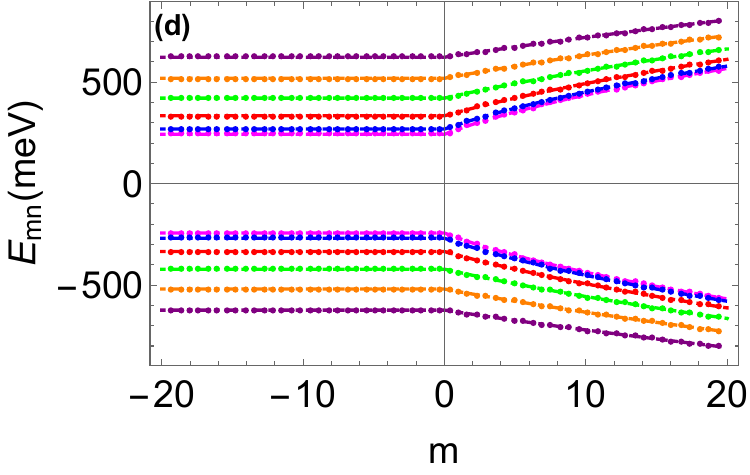}
	\caption{(color online) Energy levels $E_{mn}$ as a function of the angular momentum $m$ for $B = 10$ T, $\Delta = 0$ meV, and $R=70$ nm by varying the value of the Aharonov-Bohm flux $\phi_i$: (a, b) for $\phi_i = 0$ and (c,d) for $\phi_i = 3.5$. We choose $n = 0$ (magenta line), $n=1$ (blue line), $n=2$ (red line), $n=3$ (green line), $n=44$ (orange line), and $n=5$ (purple line). Panels (a,c) correspond to  the valley $K$ $ (s = 1)$ and  panels (b,d)  to the valley $K^\prime$ $(s = -1)$}
	\label{fig6}
\end{figure}

Fig. \ref{fig6} illustrates the energy $E_{nm}$ Eq. \eqref{19} as a function of angular momentum $m$. The curves are plotted for $R = 70$ nm, $B = 10$ T, and $\Delta = 0$ meV such as panels (a,b) display results for $\phi_i = 0$, while panels (c,d) are plotted for $\phi_i = 3.5$. In both cases, Landau levels are presented for $n = 0$ (magenta line), $1$ (blue line), $2$ (red line), $3$ (green line), $4$ (orange line), and $5$ (purple line). Here  (a,c) correspond to the $K$ valley ($s = 1$) and (b, d) to the $K^\prime$ valley ($s = -1$).
We observe that  $E_{nm}$  exhibits linearity for negative $m$ values and becomes parabolic for positive $m$ values. For $\phi_i = 0$, the gap between the conduction and valence bands is zero in the $K$ valley but is 229.4 meV in the $K^\prime$ valley. When $\phi_i$ is set to 3.5, the gap becomes non-zero, measuring 429.2 meV for the $K$ valley and 486.7 meV for the $K^\prime$ valley, surpassing the previous value of 229.4 meV. It is evident that the presence of the Aharonov-Bohm current leads either to the creation of a band-gap or an enlargement of the band-gap size.
A crucial observation is the asymmetry of the eigenvalues $E_{mn}$ concerning the sign of the quantum number $m$ and the valley $s$, indicating that $E_{mn}(m,s)\neq E_{mn}(-m,s)$ and $E_{mn}(m,s)\neq E_{mn}(m,-s)$.

\begin{figure}[ht]
	\centering
	\includegraphics[scale=0.335]{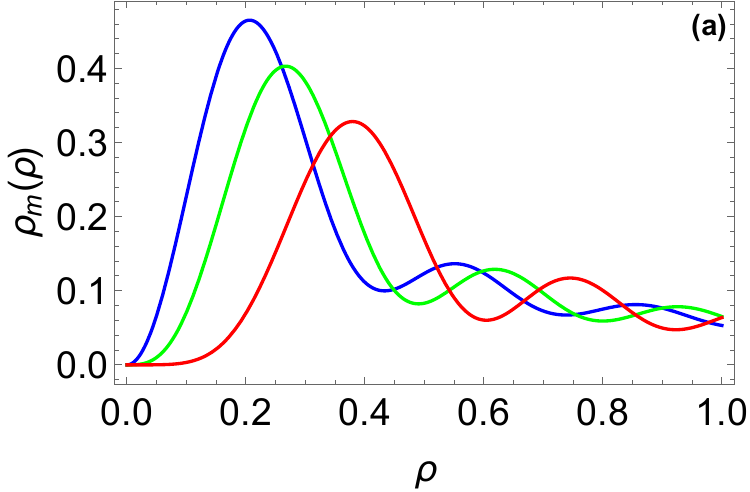}
	\includegraphics[scale=0.335]{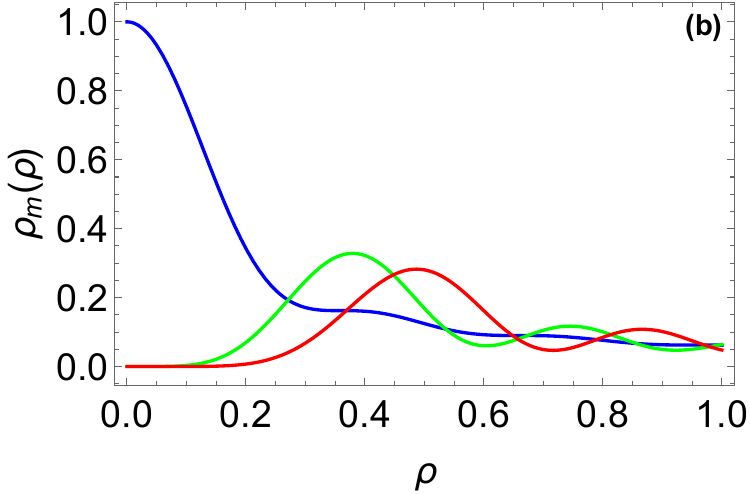}\\
	\includegraphics[scale=0.335]{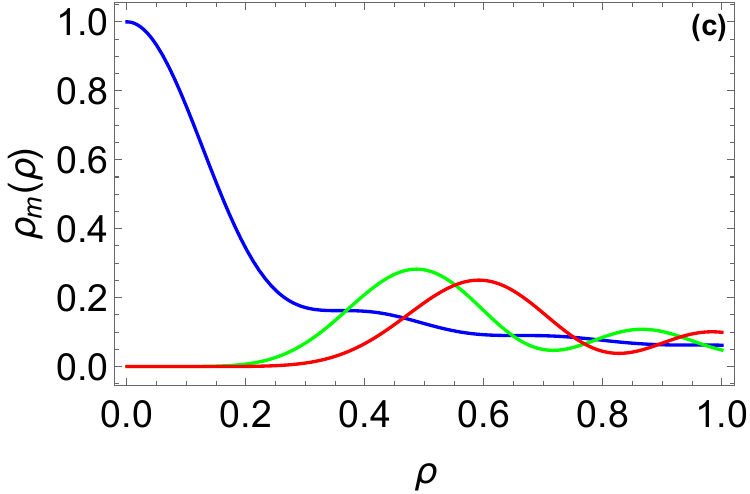}	\includegraphics[scale=0.335]{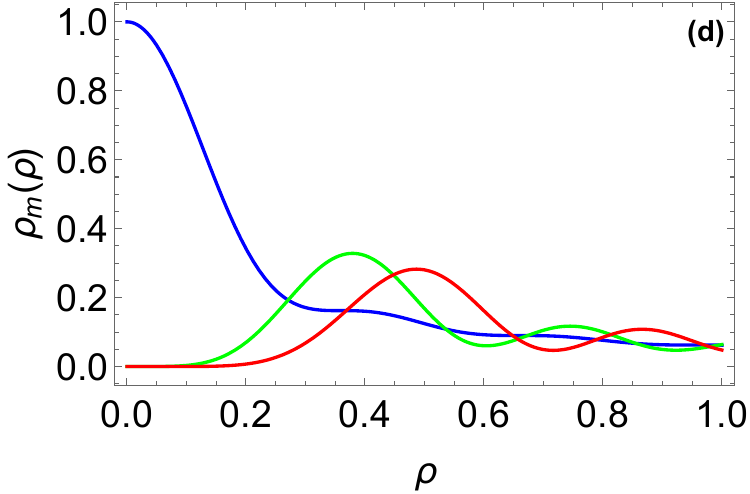}\\
	\includegraphics[scale=0.335]{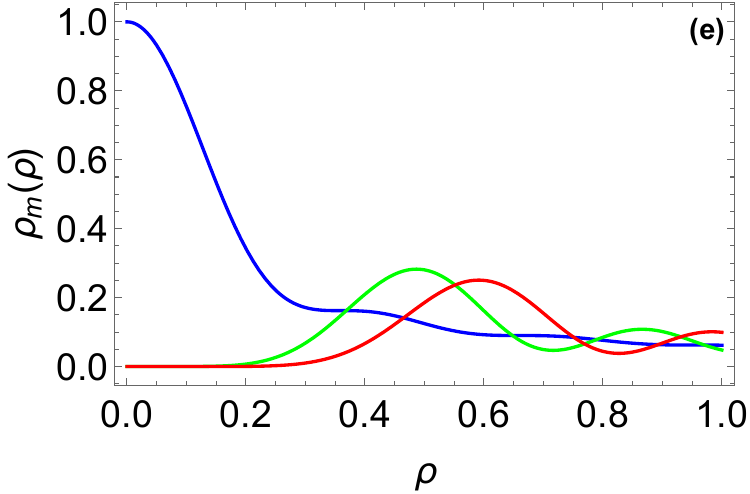}
	\includegraphics[scale=0.335]{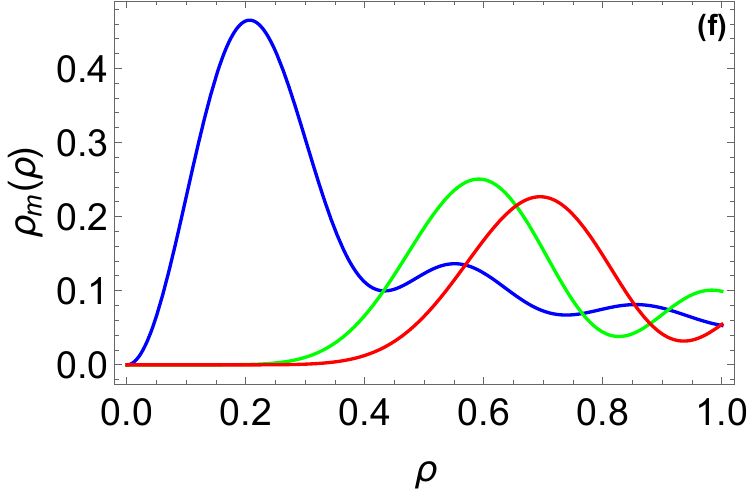}
	\caption{(color online) Radial probability $\rho_m (\rho)$ as a function of the quantum dots radius $\rho$ for $B = 10$ T and $E = 100$ meV with $m=-1$ (a,d), $m=0$ (b,e), and $m=1$ (c,f). Here, we plot by changing the values of the Aharonov-Bohm flux: $\phi_i = 0$ (blue line), 2.5 (green line), and 3.5 (red line). Panels (a,b,c) correspond to the valley $K$ ($s=1$) and panels (d,e,f) to the valley $K^\prime$ ($s=-1$).}
	\label{fig7}
\end{figure}

In Fig.~\ref{fig7}, the radial probability $\rho_m (\rho)$ is plotted against the radius of the quantum dots $\rho$ with $B=10$ T and $E=100$ meV. The conditions chosen are $m=0, \pm 1$, but with the caveat that the value of the Aharonov-Bohm flux $\phi_i$ is taken into account. The figure depicts the radial probability for three values of $\phi_i$: $\phi_i=0$ (blue line), $\phi_i=2.5$ (green line), and $\phi_i=3.5$ (red line). The top panels represent the $K$ valley ($s = 1$), while the bottom panels represent the $K^\prime$ valley when $s = -1$. Inside the quantum dot, when magnetic flux $\phi_i$ is absent (blue line), the radial probability $\rho_m(\rho)$ is observed to be maximum and equal to unity at the center of the dot ($\rho = 0$) when the product $s.m$ is positive. This occurs when $s = 1$ and $m = 0, 1$, or when $s = -1$ and $m = 0, -1$. The radial probability at the center of the quantum dot is canceled out ($\rho=0$) when the magnetic flux parameter ($\phi_i \neq 0$) is added. The highest radial probability indicates that the electron is tightly confined within the dot. It is worth noting that the inclusion of magnetic flux within the quantum dot causes a decrease in the value of this maximum by shifting it towards the edge of the quantum dot being studied. It's important to note that the radial probability shows oscillatory behavior after the first maximum. When there's magnetic flux present inside the quantum dot, there can be more than one state of electron entrapment in the dot, leading to multiple trapped electrons. Let's consider a specific scenario where $\phi_i$ equals 3.5 for $m$ equal to -1 and $s$ equal to 1, which corresponds to the $K$ valley as shown in Fig. \ref{fig7} (a). The green line in the figure represents this scenario. As we can see, there are three maximums in the graph. These maximums correspond to the values of $\rho$, which are 0.26, 0.62, and 0.93. These values indicate that there are three distinct states of an electron that can be trapped inside the quantum dot. 

Upon comparing the outcomes of our study with our previously reported findings in  \cite{bouhlal2021density, bouhlal2023zero}, several noteworthy observations can be made. We have demonstrated that when electrons are confined within a region comparable in size to their wavelength, such as quantum dots or quantum rings, various external physical parameters come into play. These parameters encompass magnetic flux, electrostatic potential, gap energy, and more. Moreover, parameters intrinsic to our system play a crucial role in determining electron confinement. Quantum nanostructures based on graphene exhibit size, deformation, and magnetic field-dependent optical properties. Magnetic flux emerges as a tunable parameter capable of controlling the transport properties of the system. The interplay of these factors underscores the rich and tunable nature of electron behavior in quantum nanostructures, offering valuable insights into their optical and transport characteristics.

\section{Conclusion} \label{concl}

Our investigation centered on the confinement of Dirac fermions within graphene quantum dots. We explored carrier confinement under the influence of an external magnetic field perpendicular to the system sheet, the internal Aharonov-Bohm flux within the quantum dot, and an energy gap. With these three physical parameters in consideration, our primary goal was to discern the impact of the Aharonov-Bohm flux on carrier confinement. To conduct our study, we solved the Dirac equation for two energy bands in close proximity to the troughs of $K$ and $K^\prime$. Analytically obtaining the eigenspinors, we employed boundary conditions to deduce an equation describing the energy levels as a function of quantum dot physical parameters—specifically, the intensity of the external magnetic field and the energy gap induced by a graphene substrate. In our numerical analysis, we directed our attention toward investigating the influence of the Aharonov-Bohm flux within the quantum dot on the energy spectrum and the radial probability of electron presence inside the quantum dot.

We conducted a thorough analysis of the numerical results for various physical parameters. Beginning with an examination of the energy spectrum and its dependence on the magnetic field  $B$, we observed that, at low $B$ values (i.e., $B \to 0$), the energy eigenvalues exhibited degeneracy. However, as the magnetic field strength increased, the energy levels became more distinct. Notably, the asymmetry $E(m, s)= E(m,-s)$ persisted, even with the inclusion of the Aharonov-Bohm (AB) flux ($\phi_i$). The energy gap widened, creating a separation between the valence and conduction bands and introducing new energy states. Introducing the AB flux alone resulted in a broader band gap, but no new energy levels emerged. In the next step, we analyzed how the energy spectrum varied with the radius of the quantum dot  $R$. Our analysis revealed that the band gap, representing the energy difference between the valence and conduction bands, diminished with an increase in the quantum dot size. This conclusion held true when considering the impact of the AB flux on the energy spectrum. Interestingly, the addition of the AB flux $\phi_i$, led to an increase in the band gap without affecting the symmetry of the system or introducing new energy levels within the energy gap $\Delta$.

We showed that for a strong magnetic field, the energy levels predicted by the boundary condition model converge toward the Landau levels of graphene. This convergence indicates that in the presence of both  magnetic field and  AB flux, the energy spectrum of the graphene quantum dot decreases. Notably, the influence of the AB flux proves to be significantly more dominant. As the AB flux increases and smooths out the constant magnetic field, the quantum dot states integrate to form Landau levels. An interesting observation is the substantial increase in the energy gap as the AB flux field rises, even without the introduction of a mass term that breaks the system's symmetry. The addition of magnetic flux enhances the probability of fermion confinement within the quantum dot by widening the energy gap. Hence, magnetic flux emerges as a crucial physical parameter capable of controlling the energy gap and regulating electron flow. It prolongs the trapping time of electrons in the quantum dot and offers a means to manipulate the energy levels of graphene quantum dots. Examining radial probability provides valuable insights into the positions of bound states and the distributions of localized electrons and holes.



\end{document}